\documentclass[aps,prb,twocolumn,showpacs,floatfix,longbibliography,10pt]{revtex4-1}
\usepackage{graphicx,amsfonts,amssymb,amsmath}
\usepackage{textcomp}
\usepackage{esint}
\usepackage[colorlinks,linktocpage,bookmarks=false,citecolor=blue,linkcolor=red,urlcolor=blue]{hyperref}

\allowdisplaybreaks

\usepackage[utf8]{inputenc}
\usepackage[T1]{fontenc}
\usepackage{lmodern}
\usepackage[dvipsnames]{xcolor}
\newlength{\textlarg}

\def\beq{\begin{equation}}
\def\eeq{\end{equation}}
\def\bleq{\begin{eqnarray}}
\def\eleq{\end{eqnarray}} 
\def\bfig{\begin{figure}}
\def\efig{\end{figure}}
\def\bline{\begin{multline}}
\def\eline{\end{multline}}
\def\bremark{\begin{quotation} \noindent \small }
\def\eremark{\end{quotation}}

\newcommand{\Tr}{{\rm Tr}} 
\newcommand{\tr}{{\rm tr}} 
 
\newcommand{\mean}[1]{\langle #1 \rangle}

\newcommand{\const}{{\rm const}}

\newcommand{\jbf}{\textbf j}

\def\eps{\epsilon}
 
\def\phibf{\boldsymbol{\phi}}
\def\varphibf{\boldsymbol{\varphi}}

\def\Lamb{\Lambda}
\def\sig{\sigma}
\def\Sig{\Sigma}

\def\half{\frac{1}{2}}

\def\third{\frac{1}{3}} 
\def\quarter{\frac{1}{4}}

\def\l{{\bf l}}

\def\p{{\bf p}} 
\def\q{{\bf q}}
\def\r{{\bf r}}

\def\x{{\bf x}}
\def\y{{\bf y}} 
 
\def\A{{\bf A}}

\def\D{{\bf D}}

\def\J{{\bf J}}

\def\nablabf{\boldsymbol{\nabla}}

\def\para{\parallel}

\def\w{\omega}
\def\wn{\omega_n}
\def\wm{\omega_m}

\def\dmu{{\partial_\mu}}
\def\dnu{{\partial_\nu}}
  
\def\dt{\partial_t} 
\def\tdt{\tilde\partial_t}
\def\dk{\partial_k}
\def\tdk{\tilde\partial_k}

\def\dtau{{\partial_\tau}}

\def\inttau{\int_0^\beta d\tau}

\def\intw{\int_{-\infty}^\infty \frac{d\w}{2\pi}}

\def\dt{\partial_t}
\def\calD{{\cal D}}

\def\calO{{\cal O}}
  
\def\calR{{\cal R}}

\def\calZ{{\cal Z}}

\def\trho{{\tilde\rho}}

\graphicspath{{./figures/}}

\DeclareMathOperator{\im}{Im}
\def\bdelta{\bar\delta} 

\def\Jmua{J^a_\mu} 
\def\DJkmua{\Delta J^a_{k,\mu}} 
\def\jmua{j^a_\mu}  
\def\Djkmua{\Delta j^a_{k,\mu}} 
\def\jnub{j^b_\nu} 
\def\Amu{A_\mu} 
 
\def\Amua{A^a_\mu} 
\def\Amub{A^b_\mu} 
\def\Amuc{A^c_\mu} 
\def\Anua{A^a_\nu} 
\def\Anub{A^b_\nu} 
\def\Anuc{A^c_\nu} 
\def\Anud{A^d_\nu} 
\def\Kmunuab{K^{ab}_{\mu\nu}}
\def\Kmunuabk{K^{ab}_{k,\mu\nu}}
\def\KmunuabLamb{K^{ab}_{\Lamb,\mu\nu}}
\def\Hmunuab{H^{ab}_{\mu\nu}}
\def\Hmunuabk{H^{ab}_{k,\mu\nu}}
\def\HmunuabLamb{H^{ab}_{\Lamb,\mu\nu}}
\def\Kparaab{K^{ab}_{\para}} 
\def\Kperpab{K^{ab}_{\perp}} 
\def\Pimunuab{\Pi^{ab}_{\mu\nu}}
\def\Piparaab{\Pi^{ab}_{\para}}

\def\sigmunuab{\sig^{ab}_{\mu\nu}}

\def\Gamzdmunuab{\Gamma^{(0,2)ab}_{\mu\nu}}

\def\Szdmunuab{S^{(0,2)ab}_{\mu\nu}} 
\def\Gamkzdmunuab{\Gamma^{(0,2)ab}_{k,\mu\nu}} 
\def\Gamuuimua{\Gamma^{(1,1)a}_{i\mu}}
\def\Gamuujmua{\Gamma^{(1,1)a}_{j\mu}}
\def\Suuimua{S^{(1,1)a}_{i\mu}}
\def\Gamkuuimua{\Gamma^{(1,1)a}_{k,i\mu}}
\def\Gamkuujmua{\Gamma^{(1,1)a}_{k,j\mu}}
\def\Gamuujmua{\Gamma^{(1,1)a}_{j\mu}}
\def\Gamuujnub{\Gamma^{(1,1)b}_{j\nu}}
\def\Sddmunuab{S^{(2,2)ab}_{\mu\nu}} 
\def\Fmunu{F_{\mu\nu}}
\def\Fmunua{F^a_{\mu\nu}}
\def\Cdis{C_{\rm dis}} 
\def\Lord{L_{\rm ord}} 
\def\GamA{\Gamma_A}
\def\GamB{\Gamma_B} 
\def\perm{\mbox{perm}} 
\def\sigA{\sig_{\rm A}}
\def\sigB{\sig_{\rm B}}

\begin{document}
\title{Nonperturbative functional renormalization-group approach to transport in the vicinity of a $(2+1)$-dimensional O($N$)-symmetric quantum critical point}

\author{F. Rose}
\author{N. Dupuis}
\affiliation{Laboratoire de Physique Th\'eorique de la Mati\`ere Condens\'ee,
CNRS UMR 7600, UPMC-Sorbonne Universit\'es, 4 Place Jussieu, 
75252 Paris Cedex 05, France}

\date{October 20, 2016} 

\begin{abstract}
Using a nonperturbative functional renormalization-group approach to the two-dimensional quantum O($N$) model, we compute the low-frequency limit $\w\to 0$ of the zero-temperature conductivity in the vicinity of the quantum critical point. Our results are obtained from a derivative expansion to second order of a scale-dependent effective action in the presence of an external (i.e., non-dynamical) non-Abelian gauge field. While in the disordered phase the conductivity tensor $\sigma(\omega)$ is diagonal, in the ordered phase it is defined, when $N\geq 3$, by two independent elements, $\sigma_{\rm A}(\w)$ and $\sigma_{\rm B}(\w)$, respectively associated to SO($N$) rotations which do and do not change the direction of the order parameter. For $N=2$, the conductivity in the ordered phase reduces to a single component $\sigma_{\rm A}(\w)$.
We show that $\lim_{\w\to 0}\sigma(\w,\delta)\sigma_{\rm A}(\w,-\delta)/\sig_q^2$ is a universal number which we compute as a function of $N$ ($\delta$ measures the distance to the quantum critical point, $q$ is the charge and $\sig_q=q^2/h$ the quantum of conductance). On the other hand we argue that the ratio $\sigma_{\rm B}(\w\to 0)/\sig_q$ is universal in the whole ordered phase, independent of $N$ and, when $N\to\infty$, equal to the universal conductivity $\sig^*/\sig_q$ at the quantum critical point.  
\end{abstract}
\pacs{05.30.Rt,74.40.Kb,05.60.Gg}

\maketitle
\tableofcontents

\section{Introduction}

Relativistic quantum field theories with O($N$) symmetry arise in the low-energy description of many condensed-matter systems: quantum antiferromagnets, superconductors, Bose-Einstein condensates in optical lattices, etc. Although the zero-temperature critical behavior at the quantum phase transition of these models as well as the finite-temperature thermodynamics\cite{Rancon13a,Rancon13b,Rancon16}
is well understood, understanding the dynamics and the transport properties in the vicinity of the quantum critical point (QCP) remains a subject of intense research.\cite{Sachdev_book} Because the corresponding theories are strongly interacting, which often leads to the absence of well-defined quasi-particles, perturbative quantum-field-theory methods are usually of limited use. So far two approaches, each with its own strengths and weaknesses, have been used: i) Quantum Monte Carlo simulations\cite{Smakov05,Chen13a,Gazit13a,Gazit14} (QMC) are a powerful method to compute imaginary-time correlation functions but suffer from the difficulty of analytically continuing data to real time; ii) holographic models\cite{Sachdev12,Witczak12,Witczak13,Katz14,Witczak14,Witczak15,Myers16,Lucas16a} based on the AdS/CFT correspondence yield real-time results but their relationship to relativistic field theories of interest in condensed matter is not always clear.

In this paper we describe a nonperturbative functional renormalization-group (NPRG) approach\cite{Berges02,Delamotte12,Kopietz_book} to the quantum O($N$) model and compute the low-frequency limit of the zero-temperature conductivity in the vicinity of the QCP. This work should be seen as a first step towards a more complete study including the frequency and temperature dependence. The NPRG has been used previously to derive the equation of state and the thermodynamic properties of the quantum O($N$) model.\cite{Rancon13a,Rancon13b,Rancon16} It has also been used to study the Higgs amplitude mode.\cite{Rancon14,Rose15}  Very good agreement with QMC simulations was obtained whenever comparison was possible. 

The method of choice to compute correlation functions from the NPRG is the Blaizot--M\'endez-Galain--Wschebor (BMW) approximation scheme since it allows one to obtain the full momentum and frequency dependence.\cite{Blaizot06,Benitez09,Benitez12} It has been used recently to compute the spectral function of the Higgs amplitude mode in the quantum O($N$) model.\cite{Rose15} Unfortunately this method violates the gauge invariance of the theory when the matter field is coupled to a gauge field and therefore cannot be used to compute the conductivity. Our results are based on a derivative expansion of the effective action, a common approximation scheme in the NPRG approach, which allows a natural implementation of symmetries.

The outline of the paper is as follows. In Sec.~\ref{sec_conductivity} we introduce the quantum O($N$) model in the presence of an external (i.e., nondynamical) non-Abelian gauge field and define the conductivity. We show how the latter can be computed in the effective action formalism. In Sec.~\ref{sec_nprg}, we introduce the scale-dependent effective action, which is the central quantity in the NPRG approach. The most general expression of the effective action to second order of the derivative expansion, compatible with gauge invariance, is defined by five functions of the (matter) field. We derive the RG equations satisfied by these functions and discuss the one-loop approximation. In Sec.~\ref{sec_largeN} we discuss the large-$N$ limit in the presence of the gauge-field-dependent regulator introduced to implement the NPRG approach. 
The numerical solution of the RG equations is presented in Sec.~\ref{sec_num}. First we briefly discuss the universal values of the critical exponents and the zero-temperature ratio $\rho_s/\Delta$ between the stiffness and the excitation gap (for two points in the phase diagram located symmetrically with respect to the QCP). Then we discuss the low-frequency limit of the conductivity. The disordered phase behaves as a capacitor and its low-frequency conductivity $\sigma(\w)=-i\w\Cdis$ is fully determined by the capacitance (per unit surface) $\Cdis$. In the ordered phase, the conductivity tensor is defined by two elements (provided that $N\geq 3$), $\sigA(\w)$ and $\sigB(\w)$, respectively associated to SO($N$) rotations which do and do not change the direction of the order parameter. (For $N=2$ the conductivity tensor reduces to $\sigA(\w)$.) $\sigA(\w)=i/\Lord(\w+i0^+)$ is the conductivity of a perfect inductor. We compute the universal ratio $\Cdis/\Lord\sig_q^2$ as a function of $N$ (with $\sig_q=q^2/h$ the quantum of conductance and $q$ the charge) and argue that the ratio $\sigB(\w\to 0)/\sig_q$ takes a universal value $\sigB^*/\sig_q$, independent of $N$ and, when $N\to\infty$, equal to the universal conductivity $\sig^*/\sig_q$ at the quantum critical point. In the Conclusion, we discuss possible improvements of our approach.

\section{Conductivity of the O($N$) model} 
\label{sec_conductivity}

The two-dimensional quantum O($N$) model is defined by the (Euclidean) action
\begin{equation}
S = \int_\x \biggl\lbrace \half (\nablabf\varphibf)^2 + \frac{1}{2c^2} (\dtau \varphibf)^2 
+ \frac{r_0}{2} \varphibf^2 + \frac{u_0}{4!N} (\varphibf^2)^2  \biggr\rbrace, 
\label{action1} 
\end{equation}
where we use the shorthand notation 
\begin{equation}
\x=(\r,\tau), \quad \int_\x = \inttau \int d^2r .
\end{equation}
$\varphibf(\x)$ is an $N$-component real field,  $\r$ a two-dimensional coordinate, $\tau\in [0,\beta]$ an imaginary time, $\beta=1/T$ and we set $\hbar=k_B=1$. $r_0$ and $u_0$ are temperature-independent coupling constants and $c$ is the (bare) velocity of the $\varphibf$ field.
The model is regularized by an ultraviolet cutoff $\Lambda$. In order to maintain the Lorentz invariance of the action~(\ref{action1}) at zero temperature, it is natural to implement a cutoff on both momenta and frequencies. 

At zero temperature there is a quantum phase transition between a disordered phase ($r_0>r_{0c}$) and an ordered phase ($r_0<r_{0c}$) where the O($N$) symmetry of the action~(\ref{action1}) is spontaneously broken (we consider $u_0$ and $c$ as fixed parameters and vary $r_0$ to induce the quantum phase transition). The QCP at $r_0=r_{0c}$ is in the universality class of the three-dimensional classical O($N$) model with a dynamical critical exponent $z=1$ (this value follows from Lorentz invariance); the phase transition is governed by the three-dimensional Wilson-Fisher fixed point. 

In the following, we consider only the zero temperature limit where the two-dimensional quantum O($N$) model is equivalent to the three-dimensional classical O($N$) model. For convenience, we set the velocity $c$ equal to one so that the action~(\ref{action1}) takes the usual form of the classical O($N$) model with $\x$ a three-dimensional space variable.\cite{no1} Having in mind the two-dimensional quantum O($N$) model, we shall refer to the critical point of the three-dimensional classical O($N$) as the QCP. In Fourier space, a correlation function $\chi(p_x,p_y,p_z)$ computed in the classical model should be identified with the correlation function $\chi(p_x,p_y,i\wn)$ in the quantum model, with $\wn\equiv p_z$ a bosonic Matsubara frequency, and yields the retarded dynamical correlation function $\chi^R(p_x,p_y,\w)\equiv \chi(p_x,p_y,i\wn\to\w+i0^+)$ after analytic continuation $i\wn\to \w+i0^+$. (At zero temperature, the Matsubara frequency $\wn=2\pi n T$ ($n$ integer) becomes a continuous variable.)  
In the following we note $d$ the space dimension of the quantum O($N$) model and $D=d+1$. Although we are primarily interested in the case $d=2$, we shall often derive expressions valid for arbitrary $d$. 

\subsection{Rotation invariance and current densities} 

The action of the $D$-dimensional O($N$) model [Eq.~(\ref{action1})] is invariant in the global rotation $\varphibf'=O\varphibf$ with $O\in\mbox{SO}(N)$ a uniform rotation. We can make this global invariance a local one by introducing a gauge field $A_\mu$ in the action:  
\begin{equation}
S = \int_\x \biggl\lbrace \half \sum_{\mu=x,y,z}(\dmu\varphibf-q\Amu\varphibf)^2  
+ \frac{r_0}{2} \varphibf^2 + \frac{u_0}{4!} (\varphibf^2)^2  \biggr\rbrace .
\label{action2} 
\end{equation}
We set the charge $q$ equal to unity in the following (it will be reintroduced in final expressions whenever necessary). $\Amu$ is a $\x$-dependent skew-symmetric matrix of the Lie algebra so($N$) (see Appendix~\ref{app_SON}); it can be written as 
\begin{equation}
\Amu = \Amua T^a 
\end{equation}
(with an implicit sum over repeated discrete indices), where $\lbrace T^a\rbrace$ denotes a set of so($N$) generators (made of $N(N-1)/2$ linearly independent skew-symmetric matrices). The action~(\ref{action2}) is invariant in the local gauge transformation, 
\begin{equation}
\begin{split} 
\varphibf' &= O\varphibf , \\
\Amu' &= O \Amu O^T +(\dmu O)O^T , 
\end{split}
\label{gaugetransform}
\end{equation}
where $O$ is a space-dependent SO($N$) rotation. Note that, {\it stricto sensu}, gauge invariance is satisfied only if it is not broken by the UV regularization. We shall assume here that this is the case and come back to this issue in Sec.~\ref{sec_nprg}. 

We do not consider $\Amu$ as a dynamical gauge field but rather as a mere external source which allows us to define the current density 
\begin{equation}
\Jmua(\x) = - \frac{\delta S}{\delta \Amua(\x)} . 
\end{equation}
To alleviate the notation we do not indicate the dependence of $\Jmua(\x)$ on $\A$. From~(\ref{action2}) we obtain 
\begin{equation}
\begin{split}
\Jmua &= \jmua -\Amu\varphibf \cdot T^a \varphibf ,  \\
\jmua &= \dmu \varphibf \cdot T^a \varphibf ,
\end{split}
\label{Jdef}
\end{equation}
where $\jmua$ and $-\Amu\varphibf \cdot T^a \varphibf$ respectively denote the ``paramagnetic'' and ``diamagnetic'' parts of the current density. Equations~(\ref{Jdef}) can also be derived from Noether's theorem. In the quantum O($N$) model, $\jbf^a=(j_x^a,j_y^a)$ corresponds to the current density (in the absence of external gauge field) whereas $j^a_z$, after a Wick rotation $z\equiv\tau\to it$, gives the conserved charge density $L^a=-ij^a_z=-\dt \varphibf\cdot T^a\varphibf$ (angular momentum density).\cite{Sachdev_book} 

For $N=2$, there is a single generator $T$, which can be chosen as minus the antisymmetric tensor $\eps_{ij}$ (Appendix~\ref{app_SON}), and we find 
\begin{equation}
J_\mu = -i [\psi^* \dmu\psi - (\dmu\psi^*)\psi] -2|\psi|^2 \Amu , 
\end{equation}
where $\psi=(\varphi_1 + i \varphi_2)/\sqrt{2}$. We thus recover the standard expression of the current density of bosons described by a complex field $\psi$.

\subsection{Linear response and conductivity} 

To leading order, the mean-value of the current density is given by 
\begin{equation}
\mean{\Jmua(\x)} = \int_{\x'} \Kmunuab(\x-\x') \Anub(\x') + \calO(A^2) , 
\end{equation}
where 
\begin{equation}
\Kmunuab(\x-\x') = \frac{\delta^2 \ln \calZ[\A]}{\delta\Amua(\x)\delta\Anub(\x')} \biggl|_{\A=0} , 
\label{Kdef} 
\end{equation}
with $\calZ[\A]$ the partition function in the presence of the external gauge field. An elementary calculation gives 
\begin{equation}
\Kmunuab(\x-\x') = \Pimunuab(\x-\x') - \delta_{\mu\nu}\delta(\x-\x')\mean{T^a\varphibf\cdot T^b\varphibf},
\end{equation}
where  
\begin{equation}
\Pimunuab(\x-\x') = \mean{\jmua(\x) \jnub(\x')} 
\end{equation} 
is the paramagnetic current-current correlation function. 

In the quantum model, the response to a uniform time-dependent gauge field is given by $\Kmunuab(i\wn)\equiv\Kmunuab(p_x=0,p_y=0,i\wn)$ and the frequency-dependent conductivity is equal to 
\begin{equation}
\sigmunuab(i\wn) = - \frac{1}{\wn} \Kmunuab(i\wn) . 
\end{equation}
The real-frequency conductivity is thus defined by 
\begin{equation}
\sigmunuab(\w) = \frac{1}{i(\w+i0^+)} \Kmunuab{}^R(\w) ,
\label{sigR}
\end{equation}
where $\Kmunuab{}^R(\w)=\Kmunuab(i\wn\to\w+i0^+)$ denotes the retarded part of $\Kmunuab(i\wn)$.

\subsection{The effective action formalism} 
\label{subsec_eaf} 

Let us consider the partition function 
\begin{equation}
\calZ[\J,\A] = \int\calD[\varphibf] \, e^{-S[\varphibf,\A] + \int_\x \J\cdot\varphibf} 
\label{ZJA}
\end{equation}
in the presence of both the gauge field $\A$ and an external source $\J$ which couples linearly to the $\varphibf$ field. The action $S[\varphibf,\A]$ is defined by~(\ref{action2}). The order parameter is obtained from 
\begin{equation}
\phibf[\x;\J,\A] = \frac{\delta\ln \calZ[\J,\A]}{\delta\J(\x)} . 
\label{phidef} 
\end{equation}
The effective action
\begin{equation}
\Gamma[\phibf,\A] = - \ln \calZ[\J,\A] + \int_\x \J\cdot\phibf 
\label{Gamdef} 
\end{equation}
is defined as the Legendre transform of $-\ln \calZ[\J,\A]$ wrt the linear source $\J$. In Eq.~(\ref{Gamdef}), $\J(\x)\equiv\J[\x;\phibf,\A]$ is a functional of $\phibf$ and $\A$ obtained by inverting~(\ref{phidef}). $\Gamma$ satisfies the equation of state
\begin{equation}
\frac{\delta\Gamma[\phibf,\A]}{\delta\phibf(\x)} = \J[\x;\phibf,\A] . 
\end{equation}

Thermodynamic properties of the system can be obtained from the effective potential 
\begin{equation}
U(\rho) = \frac{1}{V}\Gamma[\phibf,\A]\Bigl|_{\phibf=\const,\A=0}  
\end{equation}
defined by the effective action evaluated in a uniform field configuration $\phibf(\x)=\phibf$ and $\A=0$ ($V=L^D$ is the volume). For symmetry reasons, $U$ is a function of the O($N$) invariant $\rho=\phibf^2/2$. We denote by $\rho_0$ the value of $\rho$ at the minimum of the effective potential. Spontaneous symmetry breaking of the O($N$) symmetry is characterized by a nonvanishing value of $\rho_0$: $U'(\rho_0)=0$ (the prime denote derivatives wrt $\rho$).\cite{no2}

Correlation functions can be reconstructed from the one-particle irreducible (1PI) vertices defined by 
\begin{multline}
\Gamma^{(n,m)\{a_j\}}_{\{i_j\},\{\mu_j\}}[\{\x_j\},\{\y_j\};\phibf,\A] \\ = 
\frac{\delta^{n+m} \Gamma[\phibf,\A]}{\delta\phi_{i_1}(\x_1)\cdots \delta\phi_{i_n}(\x_n)\delta A_{\mu_1}^{a_1}(\y_1) \cdots \delta A_{\mu_m}^{a_m}(\y_m)} . 
\end{multline}
The correlation functions evaluated for $\A=0$ and in a uniform field configuration are determined by the vertices 
\begin{multline}
\Gamma^{(n,m)\{a_j\}}_{\{i_j\},\{\mu_j\}}(\{\x_j\},\{\y_j\};\phibf) \\ =
\Gamma^{(n,m)\{a_j\}}_{\{i_j\},\{\mu_j\}}[\{\x_j\},\{\y_j\};\phibf,\A] \Bigl|_{\phibf=\const\atop \A=0} . 
\end{multline}
In particular, the (connected) propagator $G_{ij}(\p,\phibf)=\mean{\varphi_i(\p)\varphi_j(-\p)}-\mean{\varphi_i(\p)}\mean{\varphi_j(-\p)}$  in a uniform field and for ${\A=0}$ is obtained from the matrix equation
\begin{equation}
G(\p,\phibf) = \Gamma^{(2,0)}(\p,\phibf){}^{-1} ,
\end{equation}
where $\Gamma^{(2,0)}(\p,\phibf)\equiv\Gamma^{(2,0)}(\p,-\p,\phibf)$. The O($N$) symmetry allows us to write 
\begin{equation}
\Gamma^{(2,0)}_{ij}(\p,\phibf) = \delta_{ij} \GamA(\p,\rho) + \phi_i \phi_j \GamB(\p,\rho) ,  
\label{GamAB}
\end{equation}
and 
\begin{equation}
G_{ij}(\p,\phibf) = \frac{\phi_i\phi_j}{2\rho} G_{\rm L}(\p,\rho) + \left(\delta_{i,j} - \frac{\phi_i\phi_j}{2\rho} \right) G_{\rm T}(\p,\rho) ,
\label{Gunif} 
\end{equation}
where 
\begin{equation}
\begin{split}
G_{\rm L}(\p,\rho) &= [\Gamma_{A}(\p,\rho) + 2 \rho \Gamma_{B}(\p,\rho) ]^{-1} , \\ 
G_{\rm T}(\p,\rho) &= \Gamma_{A}(\p,\rho){}^{-1} 
\end{split}
\label{GLT} 
\end{equation}
are the longitudinal (L) and transverse (T) parts of the propagator.
Due to rotation invariance in space, 2-point vertices and correlation functions in~(\ref{GamAB}) and (\ref{GLT}) are functions of $|\p|$. Since $\Gamma^{(2)}_{ij}(\p=0,\phi)=\partial_{\phi_i}\partial_{\phi_j}U(\rho)$, one has 
\begin{equation}
\GamA(\p=0,\rho) = U'(\rho) , \quad \GamB(\p=0,\rho) = U''(\rho) . 
\end{equation} 

In the disordered phase ($\rho_0=0$) the correlation length $\xi$ is finite, which corresponds to a nonzero single-particle excitation gap $\Delta\equiv c/\xi$ ($c$ is set to $1$) in the quantum model. The latter is obtained from the spectral function $\im[G^R(\q,\w,\rho=0)]$ where it manifests itself as a sharp peak. It can be more simply obtained from a derivative expansion of the 2-point vertex, 
\begin{equation}
\Gamma_A(\p,\rho) = Z(\rho) \p^2 + U'(\rho) + \calO(|\p|^4) .
\label{GamAder} 
\end{equation}
The estimate obtained from~(\ref{GamAder}), 
\begin{equation}
\Delta = \left( \frac{U'(0)}{Z(0)} \right)^{1/2} , 
\label{Deltadef} 
\end{equation}
turns out to be in very good agreement with the excitation gap deduced from the spectral function $\im[G^R(\q,\w,\rho=0)]$.\cite{Rose15}

In the ordered phase ($\rho_0>0$), the stiffness $\rho_s$ is defined by\cite{Chaikin_book} 
\begin{equation}
G_{\rm T}(\p,\rho_0) = \frac{2\rho_0}{\rho_s\p^2} \quad \mbox{for} \quad \p\to 0 . 
\end{equation}
From~(\ref{GLT},\ref{GamAder}) we deduce
\begin{equation}
\rho_s = 2 Z(\rho_0) \rho_0 .
\label{rhosdef} 
\end{equation}

\subsubsection{Conductivity from 1PI vertices} 

\begin{figure} 
\centerline{\includegraphics[width=7cm]{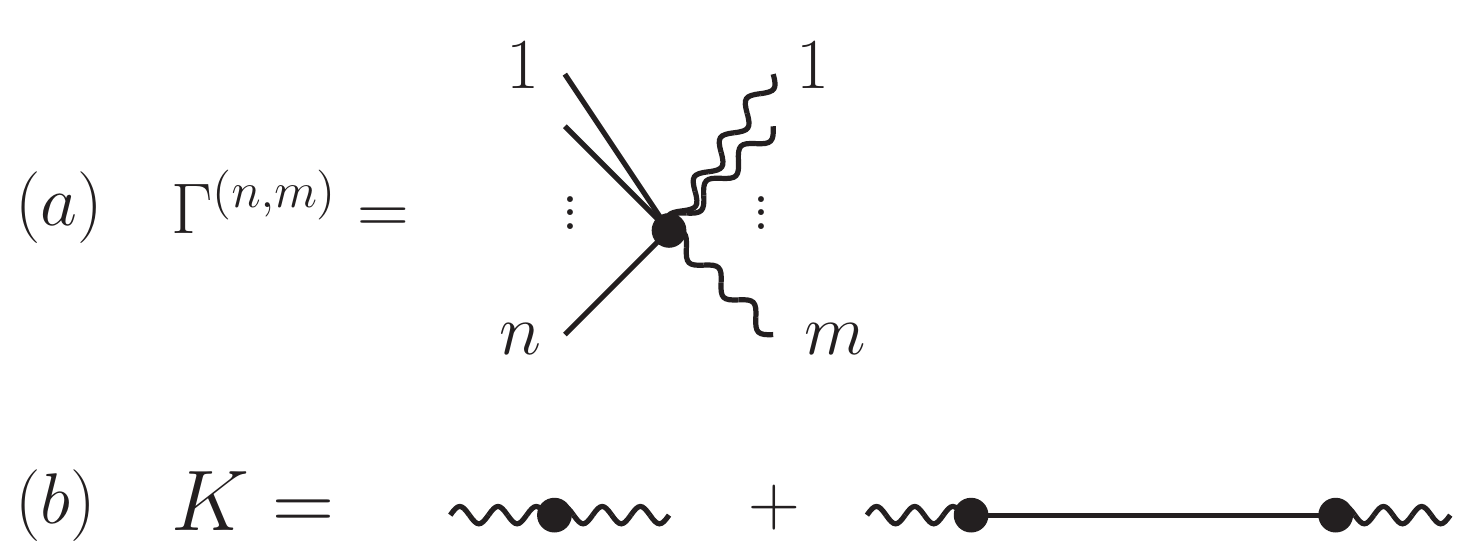}}
\caption{(a) A vertex $\Gamma^{(n,m)}$ is represented by a black dot with $n$ solid lines and $m$ wavy lines. (b) Diagrammatic representation of $\Kmunuab$ [Eq.~(\ref{K2})]. Solid lines connecting vertices stand for the propagator $G=\Gamma^{(2,0)-1}$.} 
\label{fig_K} 
\end{figure}

The conductivity can be expressed in terms of the 1PI vertices. From Eqs.~(\ref{Kdef},\ref{Gamdef}), we deduce 
\begin{align} 
\Kmunuab(\y-\y') &= \frac{\delta^2 \ln \calZ[\J,\A]}{\delta\Amua(\y)\delta\Anub(\y')} \biggl|_{\J=\A=0} \nonumber \\ 
&= - \frac{\bar\delta^2 \Gamma[\phi[\A],\A]}{\bar\delta\Amua(\y)\bar\delta\Anub(\y')} \biggl|_{\A=0} 
\label{K1}
\end{align}
where the order parameter $\phibf[\A]$ is defined by 
\begin{equation}
\frac{\delta\Gamma[\phibf,\A]}{\delta\phibf(\x)} \biggl|_{\phibf=\phibf[\A]} = 0 . 
\label{phibardef} 
\end{equation}
In Eq.~(\ref{K1}), $\bdelta/\bdelta \Amua(\y)$ is a total derivative which acts both on $\phibf[\A]$ and the explicit $\A$-dependence of the functional $\Gamma[\phibf,\A]$. In Appendix~\ref{app_K} we show that Eq.~(\ref{K1}) leads to 
\begin{multline} 
\Kmunuab(\p) = - \Gamzdmunuab(\p,\bar\phibf) \\
+ \Gamuuimua(-\p,\bar\phibf) \Gamma^{(2,0)-1}_{ij}(\p,\bar\phibf) \Gamuujnub(\p,\bar\phibf) , 
\label{K2} 
\end{multline} 
where $\bar\phibf=\phibf[\A=0]$ is the (uniform) order parameter in the absence of the gauge field and we use the notation $\Gamma^{(n,m)}(\p)\equiv\Gamma^{(n,m)}(\p,-\p)$ for both vertices $\Gamma^{(1,1)}$ and $\Gamma^{(0,2)}$. 
$\bar\phibf$ has a  modulus equal to $\sqrt{2\rho_0}$ but its direction is arbitrary. The second term in the rhs of Eq.~(\ref{K2}) corresponds to the part of $\Kmunuab$ which is not 1PI; we shall see that it does not contribute to the conductivity $\sigmunuab(\w)$ of the quantum model. Equation~(\ref{K2}) is shown diagrammatically in Fig.~\ref{fig_K}. 

As in the case of $\Gamma^{(2,0)}$ [Eq.~(\ref{GamAB})], one can take advantage of the symmetries of the model to write the vertices in the form 
\begin{equation}
\begin{split} 
\Gamuujmua(\p,\phibf) &= ip_\mu (T^a\phibf)_j \Psi_A, \\ 
\Gamzdmunuab(\p,\phibf) &= p_\mu p_\nu [ \delta_{ab} \Psi_B + (T^a\phibf)\cdot(T^b\phibf)\Psi_C ] \\ 
& + \delta_{\mu\nu} [ \delta_{ab} \bar\Psi_B + (T^a\phibf)\cdot(T^b\phibf)\bar\Psi_C ],
\end{split}
\label{vert1} 
\end{equation}
where $\Psi_A,\Psi_B,\Psi_C,\bar\Psi_B,\bar\Psi_C$ are functions of $\rho$ and $\p^2$. This leads to 
\begin{align} 
\Kparaab(\p) ={}& -\delta_{ab} [\bar\Psi_B(\p,\rho_0) + \p^2\Psi_B(\p,\rho_0)] \nonumber \\ & 
- (T^a\phibf)\cdot(T^b\phibf)  [\bar\Psi_C(\p,\rho_0) + \p^2\Psi_C(\p,\rho_0) \nonumber \\ & - \p^2 G_{\rm T}(\p,\rho_0) \Psi_A(\p,\rho_0)^2 ] 
\label{Kpara}
\end{align}
and
\begin{equation}
\Kperpab(\p) = - \delta_{ab} \bar\Psi_B(\p,\rho_0) - (T^a\bar\phibf)\cdot(T^b\bar\phibf)\bar\Psi_C(\p,\rho_0) , 
\end{equation}
where $\Kparaab$ and $\Kperpab$ are the longitudinal and transverse components of $\Kmunuab$, 
\begin{equation}
\Kmunuab(\p) = \hat p_\mu \hat p_\nu \Kparaab(\p) + (\delta_{\mu\nu} - \hat p_\mu \hat p_\nu) \Kperpab(\p) ,
\end{equation}
with $\hat \p=\p/|\p|$ a unit vector parallel to $\p$. 

Finally, the functions $\Psi_A,\Psi_B,\Psi_C,\bar\Psi_B,\bar\Psi_C$ are not independent but related by the Ward identities (see Appendix~\ref{app_wi}) 
\begin{equation}
\begin{gathered} 
\p^2 \Psi_A(\p,\rho) = \GamA(\p,\rho)-U'(\rho) , \\ 
\p^2 \Psi_B(\p,\rho) + \bar\Psi_B(\p,\rho) = 0 , \\  
\p^2 \Psi_C(\p,\rho) + \bar\Psi_C(\p,\rho) = \Psi_A(\p,\rho) .
\end{gathered}
\label{wi3} 
\end{equation}
The first equation implies $\lim_{\p\to 0}\Psi_A(\p,\rho)=Z(\rho)$ where $Z(\rho)$ is defined by~(\ref{GamAder}). Relations~(\ref{wi3}) imply that $\Kmunuab$ is transverse,
\begin{align} 
\Kparaab(\p) 
&= - (T^a\bar\phibf)\cdot(T^b\bar\phibf) \Psi_A(\p,\rho_0) U'(\rho_0) G_{\rm T}(\p,\rho_0) 
\nonumber \\ 
&= 0 ,
\end{align}
where we have used $G_{\rm T}(\p,\rho)=\GamA(\p,\rho)^{-1}$ and the fact that, depending on the phase, either $U'(\rho_0)=0$ or $\bar\phibf=0$. 
The ``$f$-sum rule''
\begin{equation}
\Kparaab(\p) = \Piparaab(\p) - \mean{(T^a\varphibf)\cdot(T^b\varphibf)} = 0 
\label{fsum} 
\end{equation}
is further discussed in Appendix~\ref{app_K:subsec_fsum}.

\subsubsection{Conductivity in the quantum model} 
\label{subsec_eaf:subsubsec_condQM}

To obtain the frequency-dependent conductivity in the quantum model, one sets $\p=(0,0,\wn)$ and $\mu,\nu=x,y$ so that $p_\mu=p_\nu=0$ and the one-particle-reducible contribution in~(\ref{K2}) vanishes, $\Kmunuab(i\wn)=-\Gamzdmunuab(i\wn,\bar\phibf)$, i.e., 
\begin{multline} 
\Kmunuab(i\wn) = \delta_{\mu\nu} \bigl\{ - \Psi_A(i\wn,\rho_0) (T^a\bar\phibf)\cdot(T^b\bar\phibf) \\
 + \wn^2 [\delta_{ab} \Psi_B(i\wn,\rho_0)  + (T^a\bar\phibf)\cdot(T^b\bar\phibf) \Psi_C(i\wn,\rho_0) ] \bigr\} , 
\end{multline} 
where we have used the Ward identities~(\ref{wi3}). 

In the disordered phase ($\rho_0=|\bar\phibf|^2/2=0$), the conductivity tensor $\sigmunuab(\w)=\delta_{\mu\nu}\delta_{ab}\sig(\w)$ is diagonal with
\begin{equation}
\sig(\w) = i\w \Psi_B(i\wn=0,\rho=0) + \calO(\w^3) .
\end{equation}
Thus, at low frequencies the disordered phase behaves as a capacitor, 
\begin{equation}
\sig(\w) = - i\w \Cdis ,
\label{sigins} 
\end{equation}
with capacitance (per unit surface) 
\begin{equation}
\Cdis = - 2\pi\hbar\sig_q \Psi_B(i\wn=0,\rho=0)
\label{Cdis}
\end{equation}
(we have restored $q$ and $\hbar$, and $\sig_q=q^2/h$ is the quantum of conductance). 

In the ordered phase, when $N \geq 3$ one must distinguish between the $N-1$ generators that act on $\bar\phibf$ (class A), i.e., such that $T^a\bar\phibf\neq 0$, and the $(N-1)(N-2)/2$ generators of class B such that $T^a\bar\phibf=0$.\cite{Podolsky11} Assuming for simplicity that the order parameter is along the first direction, i.e., $\bar\phi_i=\delta_{i,1}\sqrt{2\rho_0}$, and choosing the basis $\{T^{IJ}\}$ introduced in Eq.~\eqref{eq:basisTIJ}, one finds
\begin{multline} 
\Kmunuab(i\wn) = \delta_{\mu\nu}\delta_{ab} \{ \wn^2 \Psi_B(i\wn,\rho_0) \\
+ 2\rho_0 \delta_{a\in A} [\wn^2\Psi_C(i\wn,\rho_0) - \Psi_A(i\wn,\rho_0)] \} , 
\end{multline} 
where $\delta_{a\in A}$ is equal to unity if $T^a$ is in class A and vanishes otherwise. The conductivity tensor $\sigmunuab(\w)=\delta_{\mu\nu}\delta_{ab}\sig^{aa}(\w)$ is diagonal and defined by two independent elements, $\sigA(\w)$ for $a\in A$ and $\sigB(\w)$ for $a\in B$ (note that, for a generic basis $\{T^{a}\}$, the conductivity tensor is not necessarily diagonal).
For $N=2$ there is only one so$(N)$ generator, which belongs to class A. For $N=3$, there are three generators, which can be taken as spin-one matrices $S^1$, $S^2$ and $S^3$ (see Appendix~\ref{app_SON}). $S^2$ and $S^3$ belong to class A whereas $S^1$, the generator of rotations about the order-parameter axis, belongs to class B. Thus $K_B=\mean{j_\mu^1 j_\mu^1}$, where $j^1_\mu=-i[(\dmu\varphi_2)\varphi_3 - (\dmu\varphi_3)\varphi_2]$, is related to transverse fluctuations of the order parameter.

For class A generators, at low frequencies 
\begin{equation}
\sigA(\w) = i \frac{\rho_s}{\w+i0^+} = \rho_s \left( \pi \delta(\w) + \frac{i}{\w} \right)
\end{equation}
is characteristic of a superfluid with stiffness $\rho_s=2\rho_0\Psi_A(i\wn=0,\rho_0)=2\rho_0 Z(\rho_0)$.
The system behaves as a perfect inductor,
\begin{equation}
\sigA(\w) = \frac{i}{\Lord(\w+i0^+)} ,
\label{sigA} 
\end{equation}
with inductance 
\begin{equation}
\Lord = \frac{\hbar}{2\pi\sig_q\rho_s}
\label{Lord}
\end{equation}
(restoring $q$ and $\hbar$).  

For class B generators, 
\begin{equation}
\sigB(\w) = i(\w+i0^+) \Psi_B^R(\w,\rho_0) ,
\end{equation}
where $\Psi_B^R(\w,\rho)$ denotes the retarded part of $\Psi_B(i\wn,\rho)$. 
We will argue in Sec.~\ref{sec_num} (see also Sec.~\ref{sec_largeN} for the calculation of $\Psi_B$ in the large-$N$ limit) that in the ordered phase  $\Psi_B(i\wn,\rho)$ diverges as $1/|\wn|$ for $\wn\to 0$  so that $\sigB(\w)$ takes a finite value in the limit $\w\to 0$. 

Since $A_\mu$ enters the action in the gauge invariant combination $D_\mu=\dmu-A_\mu$, its scaling dimension at the QCP must be $[A_\mu]=[\dmu]=1$. From Eq.~(\ref{Kdef}) it follows that $[K(i\wn)]=d-1$ and $[\sig(i\wn)]=d-2$.\cite{Fisher90} For $d=2$, in the vicinity of the QCP the conductivity satisfies the scaling form\cite{Fisher90,Damle97}
\begin{equation}
\begin{split} 
\sig(\w) & = \sig_q \Sig_+\left( \frac{\w+i0^+}{\Delta}\right) , \\ 
\sig_{A,B}(\w)  & = \sig_q \Sig_-^{A,B}\left( \frac{\w+i0^+}{\Delta}\right) ,
\end{split}
\label{sigscaling}
\end{equation}
where $\Sig_\pm$ is a universal scaling function and the index $+/-$ refers to the disordered/ordered phase. In the disordered phase $\Delta$ is the excitation gap. In the ordered phase, we choose $\Delta$ to be given by the excitation gap at the point of the disordered phase located symmetrically wrt the QCP (i.e., corresponding to the same value of $|r_0-r_{0c}|$), the ratio $\rho_s/\Delta$ being universal. At the QCP, the universal scaling functions reach a nonzero limit $\Sig_\pm(\infty)$ and the ratio $\sig(\w=0)/\sig_q=\Sig_\pm(\infty)$ is universal.\cite{Fisher90}

Comparing~(\ref{sigscaling}) with~(\ref{sigins}), we see that $\Sig_+(x)$ must vanish linearly with $x$ in the limit $x\to 0$ and 
\begin{equation} 
\Cdis = i \frac{\sig_q}{\Delta} \Sig'_+(0)  . 
\label{CSig} 
\end{equation}
On the other hand, Eq.~(\ref{sigA}) shows that $\Sig_-^A(x)\sim 1/x$ for $x\to 0$ and 
\begin{equation}
\frac{1}{\Lord} = -i \Delta \sig_q \lim_{x\to 0} x \Sig_-^A(x) . 
\label{LSig} 
\end{equation} 
From~(\ref{CSig}) and (\ref{LSig}) we deduce that the ratio 
\begin{align}
\frac{\Cdis}{\Lord \sig_q^2} &= \Sig'_+(0) \lim_{x\to 0} x \Sig_-^A(x) \nonumber \\ 
&= \lim_{\w\to 0} \frac{\sig(\w,\delta) \sigA(\w,-\delta)}{\sig_q^2} 
\end{align} 
is universal in the critical regime ($\delta=r_0-r_{0c}$). 

The divergence of $\Psi_B(i\wn,\rho)$ as $1/|\wn|$ for $\wn\to 0$  implies that $\Sig_-^B(x)$ has a finite limit when $x\to 0$. As a result $\sigB(\w\to 0)/\sig_q$ is a universal number in the whole ordered phase (see Sec.~\ref{sec_num}).

\section{Nonperturbative renormalization group} 
\label{sec_nprg} 

In this section we show how the NPRG allows us to compute the effective action $\Gamma[\phibf,\A]$. 

\subsection{Scale-dependent effective action}  

The strategy of the NPRG approach is to build a family of models indexed by a momentum scale $k$ such that fluctuations
are smoothly taken into account as $k$ is lowered from the microscopic scale $\Lambda$ down to 0.\cite{Berges02,Delamotte12,Kopietz_book} This is achieved by adding to the action~(\ref{action2}) the infrared regulator term
\begin{align} 
\Delta S_k[\varphibf,\A] &= \half \int_\x \varphibf(\x) \cdot R_k\bigl(-\D(\x)^2\bigr) \varphibf(\x) \nonumber \\ 
&= \half \int_{\x,\x'} \varphibf(\x) \cdot \calR_k[\x,\x',\A] \varphibf(\x') , 
\label{DeltaSk}
\end{align} 
where 
\begin{equation}
\calR_k[\x,\x',\A] = \half \Bigl[ R_k\bigl(-\D(\x)^2\bigr) + R_k\bigl(-\D(\x')^2\bigr)^T \Bigr] \delta(\x-\x') 
\end{equation} 
and $\D^2=D_\mu D_\mu$ with $D_\mu(\x)=\partial_{x_\mu}-A_\mu(\x)$. In the absence of the external gauge field, we recover the usual regulator term. By replacing $\dmu$ by the covariant derivative $D_\mu$ in the regulator term, we ensure that the action $S+\Delta S_k$ remains gauge invariant. Gauge-invariant regulators have been considered before in the context of gauge theories.\cite{Morris00,Morris00a,Arnone02,Bartosch13,Codello16} 

The regulator has to satisfy some general properties. The cutoff function $R_k$ must be of order $k^2$ for
$|\q|\ll k$ and small wrt $k^2$ for $|\q|\gg k$. That way, fluctuations with momentum
$|\q|\lesssim k$ are suppressed but those with $|\q|\gtrsim k$ are left unaffected. The exact shape of the regulator is further discussed in Sec.~\ref{sec_num}.

The partition function 
\begin{equation}
\calZ_k[\J,\A] = \int\calD[\varphibf]\, e^{-S[\varphibf,\A]-\Delta S_k[\varphibf,\A]+\int_\x \J\cdot\varphibf}  
\end{equation} 
is now $k$ dependent. The scale-dependent effective action 
\begin{equation}
\Gamma_k[\phibf,\A] = -\ln \calZ_k[\J,\A] + \int_\x \J\cdot\phibf -\Delta S_k[\phibf,\A] 
\end{equation}
is defined as a modified Legendre transform of $-\ln \calZ_k[\J,\A]$ which includes the subtraction of $\Delta S_k[\phibf,\A]$. Here $\phibf(\x)=\mean{\varphibf(\x)}$ is the order parameter (in the presence of the external source $\J$ and the gauge field $\A$).
Assuming that fluctuations are completely frozen by the $\Delta S_k$ term when $k=\Lambda$, $\Gamma_\Lambda[\phibf,\A]=S[\phibf,\A]$. On the other hand, the effective action of the original model~(\ref{action2}) is given by $\Gamma_{k=0}$ provided that $R_{k=0}$ vanishes. 

The variation of the effective action with $k$ is given by Wetterich's equation\cite{Wetterich93} 
\begin{equation}
\dt \Gamma_k[\phibf,\A] = \half \Tr \Bigl\{ \dot \calR_k[\A] \bigl(\Gamma_k^{(2,0)}[\phibf,\A] + \calR_k[\A] \bigr)^{-1} \Bigr\} ,
\label{Weteq}
\end{equation}
where $\dot \calR_k[\A]=\dt \calR_k[\A]$ and $t=\ln(k/\Lamb)$ is a (negative) RG ``time''. In Fourier space, the trace involves a sum over momenta as well as the O($N$) index of the $\phibf$ field. The regulator $\calR$ in Eq.~(\ref{Weteq}) ensures that high momenta do not contribute to the flow and the momentum integrals can be safely extended up to infinity. Thus the regulator term $\Delta S_k$ provides us with a gauge-invariant UV regularization (see also Sec.~\ref{subsec_oneloop}).  

All properties of the effective action discussed in Sec.~\ref{subsec_eaf} carry over to $\Gamma_k[\phibf,\A]$, the only change being that all quantities become $k$ dependent. Thus, the computation of the conductivity requires to determine the vertices $\Gamma^{(n,m)}_{k=0}$ for $(n,m)=(1,1)$, $(0,2)$ in addition to the effective potential $U_{k=0}$ and the inverse propagator $\Gamma^{(2,0)}_{k=0}$. 

\subsection{Derivative expansion} 
\label{sec_nprg:subsec_de}

To solve Wetterich's equation, we use a derivative expansion of the scale-dependent effective action. Such an expansion is made possible by the regulator term $\Delta S_k$ which ensures that all vertices $\Gamma^{(n,m)}_k$ are smooth functions of momenta $\p_i$ and can be expanded in powers of $\p^2_i/k^2$ when $|\p_i|\ll k$, even at criticality. Thus the derivative expansion of the effective action is justified as long as we are interested only in the long-wavelength physics (corresponding to length scales larger than $k^{-1}$ or the correlation length of the theory). It does not allow us to obtain the full frequency dependence of the conductivity but is sufficient to determine the low-frequency limit of $\sig(\w)$ and $\sigA(\w)$ defined by $\Cdis$ and $\Lord$ [Eqs.~(\ref{sigins},\ref{sigA})]. The case of $\sigB(\w)$ is more subtle; nevertheless, we shall see that the derivative expansion is sufficient to infer that the ratio $\sigB(\w\to 0)/\sig_q$ is universal in the whole ordered phase, altough the universal value $\sigB^*/\sig_q$ cannot be definitely determined. The derivative expansion does not grant access to the universal critical conductivity $\sig^*$ either. On the other hand, the derivative expansion allows a simple implementation of gauge invariance. This should be contrasted with the Blaizot--M\'endez-Galain--Wschebor approximation,\cite{Blaizot06,Benitez09,Benitez12} recently used to compute the Higgs (scalar) susceptibility,\cite{Rose15} which enables to calculate the full momentum/frequency dependence of correlation functions but violates gauge invariance. 

The derivative expansion of $\Gamma_k[\phibf,\A]$ is fully determined by the symmetries of the system. For instance, for $\A=0$, the most general O($N$) invariant effective action to second order in derivatives, 
\begin{equation}
\Gamma_k[\phibf] = \int_\x \biggl\lbrace \half Z_k(\rho) (\nablabf\phibf)^2 + \quarter Y_k(\rho)(\nablabf\rho)^2 
+ U_k(\rho) 
\biggr\rbrace 
\label{Gamder}
\end{equation}
is defined by three functions: the effective potential $U_k(\rho)$, $Z_k(\rho)$ and $Y_k(\rho)$.\cite{Berges02,Delamotte12} There are two $\calO(\nablabf^2)$ terms, reflecting the fact that longitudinal and transverse fluctuations (wrt the local order parameter $\phibf(\r)$) have different stiffness. 
The excitation gap in the disordered phase and the stiffness in the ordered phase are then defined by $\Delta_k=(W_k(0)/Z_k)^{1/2}$ and $\rho_{s,k}=2Z_k\rho_{0,k}$, respectively [see Eqs.~(\ref{Deltadef}) and (\ref{rhosdef})].

When $\A\neq 0$, the effective action must be invariant in the gauge transformation~(\ref{gaugetransform}): $\Gamma_k[\phibf,\A]=\Gamma_k[\phibf',\A']$. The effective action~(\ref{Gamder}) can be made gauge invariant by replacing $\dmu$ by the covariant derivative $D_\mu$. It may also include terms depending on the field strength\cite{no3} 
\begin{equation}
F_{\mu\nu} = - [D_\mu,D_\nu] = \dmu A_\nu -\dnu A_\mu - [A_\mu,A_\nu] . 
\end{equation}
Although $F_{\mu\nu}$ is not gauge invariant (it transforms as $F'_{\mu\nu}=OF_{\mu\nu}O^T$), it allows us to construct two invariant terms, namely $\tr(\Fmunu^2)$ and $(\Fmunu\phibf)^2$. Here $\tr$ denotes the trace wrt the O($N$) indices. This leads to the effective action 
\begin{multline} 
\Gamma_k[\phibf,\A] = \int_\x \biggl\lbrace \half Z_k(\rho) D_\mu\phibf \cdot D_\mu\phibf   + \quarter Y_k(\rho)(\nablabf\rho)^2 
+ U_k(\rho) \\ + \quarter X_{1,k}(\rho) \Fmunua\Fmunua  + \quarter X_{2,k}(\rho) \sum_{\mu\nu}(\Fmunua T^a\phibf)^2 \biggr\rbrace , 
\label{Gamder1} 
\end{multline} 
where we use $\Fmunu=\Fmunua T^a$ and $\tr(\Fmunu^2)=-2\Fmunua\Fmunua$ with
\begin{equation}
F^a_{\mu\nu} = \dmu \Anua -\dnu \Amua - f_{abc} \Amub \Anuc 
\end{equation} 
(the structure constants $f_{abc}$ of the so($N$) Lie algebra are defined in Appendix~\ref{app_SON}). Note that, as the gauge field is treated on the same footing as the gradient term, we restrict ourselves to terms of second order in $\A$.
The effective action $\Gamma_k[\phibf,\A]$ is determined by $X_{1,k}(\rho)$ and $X_{2,k}(\rho)$ in addition to $U_k(\rho)$, $Z_k(\rho)$ and $Y_k(\rho)$. From~(\ref{Gamder1}) we obtain the vertices 
\begin{align}
\Gamma_{A,k}(\p,\rho) &= Z_k(\rho) \p^2 + U'_k(\rho) , \nonumber \\*
\Gamma_{B,k}(\p,\rho) &= \frac{\rho}{2} Y_k(\rho) \p^2 + U''_k , \label{Gam11etc} \\*
\Gamkuujmua(\p,\phibf) &= ip_\mu (T^a\phibf)_j Z_k(\rho) \nonumber
\end{align} 
and
\begin{multline}
\Gamkzdmunuab(\p,\phibf) = Z_k(\rho) \delta_{\mu\nu} (T^a\phibf)\cdot(T^b\phibf)  
+ [ X_{1,k}(\rho) \delta_{ab} \\ + X_{2,k}(\rho) (T^a\phibf)\cdot(T^b\phibf) ] (\delta_{\mu\nu}\p^2 - p_\mu p_\nu) . 
\label{Gam02} 
\end{multline}
Comparing with~(\ref{vert1}) we deduce that 
\begin{equation}
\begin{gathered}
\Psi_{A,k}(\p,\rho) = Z_k(\rho) , \quad \Psi_{B,k}(\p,\rho) = - X_{1,k}(\rho), \\  
\Psi_{C,k}(\p,\rho) = - X_{2,k}(\rho), \quad \bar\Psi_{B,k}(\rho) = \p^2 X_{1,k}(\rho) , \\ 
\bar\Psi_{C,k}(\rho) = Z_k(\rho) + \p^2 X_{2,k}(\rho)  
\end{gathered}
\label{vert2} 
\end{equation}
to lowest order of the derivative expansion. Equations~(\ref{vert2}) satisfy the Ward identities~(\ref{wi3}).

\subsection{Flow equations} 

From Wetterich's equation~(\ref{Weteq}), we deduce the flow equations satisfied by the effective potential,
\begin{equation}
\dt U_k(\rho) = \half \int_\q \dot R_k(\q) [ G_{k,\rm L}(\q,\rho) + (N-1) G_{k,\rm T}(\q,\rho) ] 
\label{dUdef}
\end{equation}
($\dot R_k=\dt R_k$), and the vertices of interest in a uniform field, 
\begin{widetext} 
\begin{equation}
\begin{split} 
\dt \Gamma^{(2,0)}_{k,ij}(\p) ={}& \half \tdt \sum_\q \tr\Bigl\{ G_k(\q) \bigl[ \Gamma^{(4,0)}_{k,ij}(\p,-\p,\q,-\q) 
- \Gamma^{(3,0)}_{k,i}(\p,\q,-\p-\q) G_k(\p+\q) \Gamma^{(3,0)}_{k,j}(-\p,\p+\q,-\q) \bigr] \Bigl\} , \\ 
\dt \Gamkuuimua(\p) ={}&  \half \tdt \sum_\q \tr \Bigl\{ G_k(\q) \Gamma^{(3,1)a}_{k,i\mu}(\q,-\q,\p,-\p) \\ & 
- G_k(\q) \Gamma^{(3,0)}_{k,i}(\p,\q,-\p-\q) G_k(\p+\q) \bigl[ \Gamma^{(2,1)a}_{k,\mu}(\p+\q,-\q,-\p) + \calR^{(1)a}_{k,\mu}(\p+\q,-\q,-\p) \bigr] \Bigr\} ,
\\ 
\dt \Gamkzdmunuab(\p) ={}& \half \tdt \sum_\q \tr \Bigl\{ 
G_k(\q) \bigl[ \Gamma^{(2,2)ab}_{k,\mu\nu}(\q,-\q,\p,-\p) + \calR^{(2)ab}_{k,\mu\nu}(\q,-\q,\p,-\p) \bigr] 
 \\ & 
- G_k(\q) \bigl[ \Gamma^{(2,1)a}_{k,\mu}(\q,-\p-\q,\p) + \calR^{(1)a}_{k,\mu}(\q,-\p-\q,\p) \bigr] G_k(\p+\q) \\ & \times 
\bigl[ \Gamma^{(2,1)b}_{k,\nu}(\p+\q,-\q,-\p) + \calR^{(1)b}_{k,\nu}(\p+\q,-\q,-\p) \bigr] \Bigr\},
\end{split}
\label{rgeq} 
\end{equation} 
\end{widetext}
where $\Gamma^{(2,0)}_k(\p)\equiv\Gamma^{(2,0)}_k(\p,-\p)$, etc. To alleviate the notation we do not write explicitly the dependence of the vertices on the uniform field $\phibf$. The propagator $G_k=(\Gamma^{(2,0)}_k+R_k)^{-1}$ in~(\ref{rgeq}) includes the cutoff function and the operator $\tdt=(\dt R_k)\partial_{R_k}$ acts only on the $k$ dependence of the cutoff function $R_k$, i.e., 
\begin{equation}
\tdt G_{k,i_1i_2}(\q) = - \dot R_k(\q) G_{k,i_1i_3}(\q) G_{k,i_3i_2}(\q).
\end{equation}
Since the effective potential enters the propagators only through its derivatives, it is convenient to consider $W_k(\rho)=U'_k(\rho)$ whose flow equation can be easily deduced from~(\ref{dUdef}).

\begin{figure}
\centerline{\includegraphics[bb=8 10 750 360,width=8.5cm]{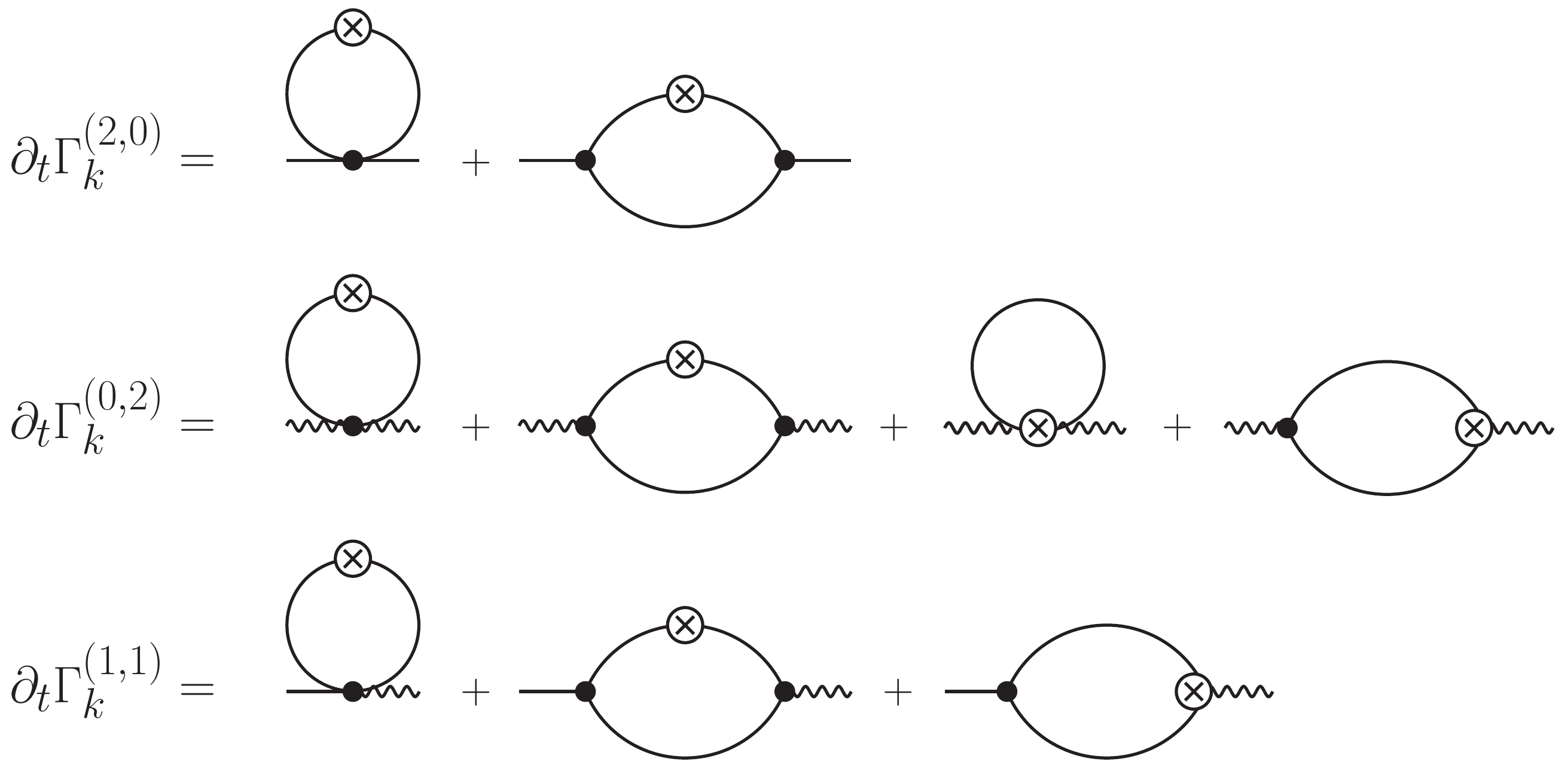}}
\caption{Diagrammatic representation of the RG equations~(\ref{rgeq}). Signs and symmetry factors are not shown. The diagrammatic representation of the vertices is shown in Fig.~\ref{fig_K} and the cross with $n$ wavy lines stands for $\dt R_k$, $\dt\calR^{(1)}_k$ and $\dt\calR^{(2)}_k$ for $n=0,1,2$, respectively.} 
\label{fig_rgeq}
\end{figure}

Equations~(\ref{rgeq}) are shown diagrammatically in Fig.~\ref{fig_rgeq}. They differ from usual flow equations by the appearance of derivatives of the cutoff function wrt the gauge field,
\begin{equation}
\begin{split} 
\calR^{(1)a}_{k,\mu}(\x,\x',\y) &= \frac{\delta \calR[\x,\x',\A]}{\delta \Amua(\y)} \biggl|_{\A=0} , \\ 
\calR^{(2)ab}_{k,\mu\nu}(\x,\x',\y,\y') &= \frac{\delta^2 \calR[\x,\x',\A]}{\delta \Amua(\y) \delta\Anub(\y')} \biggl|_{\A=0} . 
\end{split}
\label{Rcal} 
\end{equation}
The explicit expressions of $\calR^{(1)a}_{k,\mu}(\q,-\p-\q,\p)$ and $\calR^{(2)ab}_{k,\mu\nu}(\q,-\q,\p,-\p)$ are given in Appendix~\ref{app_Rcal}. 

The flow equations of the functions $Z_k(\rho)$, $Y_k(\rho)$, $X_{1,k}(\rho)$ and $X_{2,k}(\rho)$ can be deduced from~(\ref{rgeq}) using Eqs.~(\ref{Gam11etc},\ref{Gam02}). They are too long to be shown here but can be found in Appendix~\ref{app_rgeq}. Since the flow equations do not generate terms that are not already present in the scale-dependent effective action, we verify {\it a posteriori} that Eq.~(\ref{Gamder1}) gives the most general expression of $\Gamma_k$ to second order in derivatives. This in turn justifies the expression of $\Gamma^{(1,1)}$ and $\Gamma^{(0,2)}$ in Eqs.~(\ref{vert1}). 

Before discussing the numerical solution of the flow equations (Sec.~\ref{sec_num}), we consider two limiting cases where known results can be recovered and elementary checks made: the one-loop approximation and the large-$N$ limit.  

\subsection{One-loop approximation}
\label{subsec_oneloop}

The one-loop correction to the bare effective action $\Gamma_\Lamb[\phibf,\A]=S[\phibf,\A]$ can be retrieved from the exact flow equation~(\ref{Weteq}) by replacing $\Gamma^{(2,0)}_k$ by $\Gamma^{(2,0)}_\Lamb=S^{(2,0)}$ in the rhs, i.e., 
\begin{equation}
\dt \Gamma_k[\phibf,\A] = \half \Tr \Bigl\{ \dot \calR_k[\A] \bigl(S^{(2,0)}[\phibf,\A] + \calR_k[\A] \bigr)^{-1} \Bigr\} .
\end{equation}
Integrating this expression between $k=\Lamb$ and $k=0$ with boundary condition $\Gamma_\Lamb[\phibf,\A]=S[\phibf,\A]$  we obtain the effective action to one-loop order, 
\begin{align}
\Gamma[\phibf,\A] ={}& S[\phibf,\A] + \half \Tr \bigl\{ \ln(S^{(2,0)}[\phibf,\A]) \nonumber \\ & 
- \ln(S^{(2,0)}[\phibf,\A]+\calR_\Lamb[\A]) \bigr\} .
\label{Gam1loop}
\end{align} 
The advantage of this expression over the usual one-loop expression is that the regulator $\calR_\Lamb[\A]$ is used as a gauge-invariant UV cutoff and the momentum integrals can be extended up to infinity. In Appendix~\ref{app_1loop}, we show that the three expressions of $Z(\rho)$, obtained from $\Gamma^{(2,0)}$, $\Gamma^{(1,1)}$ and $\Gamma^{(0,2)}$, are in agreement.

\section{The large-$N$ limit} 
\label{sec_largeN}

In this section we compute the conductivity in the large-$N$ limit within the standard approach\cite{Sachdev_book} but from the action $S+\Delta S_k$ including the gauge-field-dependent regulator. We then discuss to what extent the results can be recovered from the flow equations. 

\subsection{Conductivity}
\label{sec_largeN:subsec_direct}

The regulator term $\Delta S_k$ gives the additional contribution~(\ref{DJmuadef})
to the current density~(\ref{Jdef}). Since we are interested in the linear response, we need $\DJkmua$ to first order in $\A$, 
\begin{multline} 
\DJkmua(\x) = - \half \int_{\y,\y'} \varphibf(\y) \cdot \calR^{(1)a}_{k,\mu}(\y,\y',\x) \varphibf(\y') \\
- \half \int_{\y,\y',\x'} \varphibf(\y) \cdot \calR^{(2)ab}_{k,\mu\nu}(\y,\y',\x,\x') \varphibf(\y') \Anub(\x') . 
\end{multline}
This modifies the linear response function $\Kmunuab$ in two ways. 

On the one hand the paramagnetic current-current correlation function should now be computed with the paramagnetic current (in Fourier space) 
\begin{multline}
\jmua(\p)+\Djkmua(\p) = \frac{1}{\sqrt{V}} \sum_{\q} \varphibf(\q) \cdot \Bigl[ iq_\mu T^a  \\
 - \half \calR^{(1)a}_{k,\mu}(-\q,\q-\p,\p) \Bigr] \varphibf(\p-\q) . 
\end{multline} 
In the quantum model, assuming a cutoff function $R_k$ which acts only on momenta and using $\calR^{(1)a}_{k,\mu}(-\q,\q,0)=-2i q_\mu R_k'(\q^2)T^a$ (Appendix~\ref{app_Rcal}), we obtain 
\begin{multline} 
\jmua(i\wm)+\Djkmua(i\wm) = \frac{1}{\beta L^d} \sum_{\q,\wn}  iq_\mu(1+R_k') \\ \times \varphibf(\q,i\wn)\cdot 
T^a \varphibf(-\q,i\wm-i\wn) 
\end{multline} 
where $\mu\in\{x,y\}$, $R'_k\equiv R'_k(\q^2)$, and $L^d$ is the volume. We find 
\begin{align} 
\Pimunuab(i\wm) ={}& 2 \delta_{ab}  \sum_{i,j} T^a_{ij}{}^2 T \sum_{\wn} \int_\q q_\mu q_\nu (1+R'_k)^2 \nonumber \\ 
& \times G_i(\q,i\wn) G_j(\q,i\wn+i\wm) ,
\label{largeN1}
\end{align} 
to leading order in the large-$N$ limit. For simplicity we assume that the order parameter lies along a particular direction, i.e., $\mean{\varphi_i}=\delta_{i,i_0} \mean{\varphi_{i_0}}=\delta_{i,i_0}\sqrt{2\rho_0}$. This way, the propagator is diagonal wrt the O($N$) indices and $G_{ii}\equiv G_i=G_T+\delta_{i,i_0}(G_L-G_T)$.

On the other hand there is an additional diamagnetic contribution to $\Kmunuab(\p)$, 
\begin{equation}
- \frac{1}{2} \int_\q \mean{ \varphibf(\q) \cdot \calR^{(2)ab}_{k,\mu\nu}(-\q,\q,\p,-\p) \varphibf(-\q) } .
\end{equation}
Using 
\beq
\calR^{(2)ab}_{k,\mu\nu}(-\q,\q,0,0) = - \{T^a,T^b\} [ \delta_{\mu\nu} R'_k + 2 q_\mu q_\nu R''_k ] 
\eeq
(see Appendix~\ref{app_Rcal}), we obtain the total diamagnetic contribution to $\Kmunuab(i\wm)$, 
\begin{align} 
& - \delta_{\mu\nu} \delta_{ab} \sum_{i,j} T^a_{ij}{}^2 T \sum_{\wn} \int_{\q} \Bigl( 1 + R'_k + \frac{2}{d} \q^2 R''_k \Bigr) \nonumber \\ 
& \times [ G_j(\q,i\wn) + \mean{\varphi_j(\q,i\wn)} \mean{\varphi_j(-\q,-i\wn)} ].
\label{largeN2}
\end{align} 
 
The momentum integrals in Eqs.~(\ref{largeN1}) and (\ref{largeN2}) are not convergent and a UV cutoff is necessary, which violates gauge invariance. To circumvent this difficulty, we notice that the large-$N$ expression $\Kmunuabk=\Hmunuabk$ of the response function $\Kmunuab$ can be rewritten as a RG equation $\dk\Kmunuabk=\dk\Hmunuabk$, where $\Hmunuabk$ is defined as the sum of~(\ref{largeN1}) and (\ref{largeN2}). No upper cutoff is necessary when computing $\dk\Hmunuabk$ since the UV convergence of the momentum integrals is ensured by the cutoff function $R_k$ and its derivatives. Integrating the RG equation between $k=0$ and $k=\Lamb$, we obtain the gauge invariant response function 
\beq
\Kmunuab(i\wm) = \KmunuabLamb(i\wm) + \Hmunuab(i\wm) -\HmunuabLamb(i\wm) , 
\label{largeN3}
\eeq
where $\KmunuabLamb(i\wm)=-\delta_{\mu\nu}\delta_{ab}\delta_{a\in A}2\rho_{0,\Lamb} (1+R'_\Lamb(0)) $ if we assume that the initial (gauge-invariant) condition of the RG equation is given by the mean-field solution: $\KmunuabLamb=-\delta^2 (S+\Delta S_\Lamb)/\delta\Amua\delta\Anub$. The term $\HmunuabLamb$ in~(\ref{largeN3}) is crucial since it guarantees that all momentum integrals are UV convergent. Equation~(\ref{largeN3}) should be compared to the gauge-invariant one-loop expression of the effective action [Eq.~(\ref{Gam1loop})]. 

This yields the final expression of $\Kmunuab=\delta_{\mu\nu}\delta_{ab}K^{aa}$ in the large-$N$ limit, 
\begin{widetext}
\begin{align} 
K^{aa}(i\wm) ={}& \sum_{i,j} T^a_{ij}{}^2 T \sum_{\wn} \int_\q \biggl\{ \frac{2}{d} \q^2  [ G_i(\q,i\wn) G_j(\q,i\wn+i\wm) - (1+R'_\Lamb)^2 G_{\Lamb,i}(\q,i\wn) G_{\Lamb,j}(\q,i\wn+i\wm) ] \nonumber \\ &
- G_j(\q,i\wn) - \mean{\varphi_j(\q,i\wn)} \mean{\varphi_j(-\q,-i\wn)} 
+ \Bigl( 1+R'_\Lamb+\frac{2}{d}\q^2 R''_\Lamb \Bigr) G_{\Lamb,j}(\q,i\wn) \biggr\}  ,
\end{align}
where $\q$ is a $d$-dimensional vector.

\subsubsection{Disordered phase} 

In the disordered phase $K^{aa}(i\wm)\equiv K(i\wm)$ does not depend on $a$, 
\begin{align} 
K(i\wm) ={}& 2 T \sum_{\wn} \int_\q \biggl\{ \frac{2}{d} \q^2  [ G(\q,i\wn) G(\q,i\wn+i\wm) - (1+R'_\Lamb)^2 G_{\Lamb}(\q,i\wn) G_{\Lamb}(\q,i\wn+i\wm) ] \nonumber \\ & 
- G(\q,i\wn) + \Bigl(1+R'_\Lamb+\frac{2}{d}\q^2 R''_\Lamb \Bigr) G_{\Lamb}(\q,i\wn) \biggr\} ,
\label{largeN4} 
\end{align}
where $G_k(\q,i\wn)=(\wn^2+\q^2+R_k+\Delta^2_k)^{-1}$ is the large-$N$ limit of the propagator\cite{Rose15} (with $G\equiv G_{k=0}$) and $\Delta^2_k=W_k(\rho=0)$ is the square of the excitation gap. 
Using the relation 
\beq
\int_\q  \q^2 (1+R'_k)^2 G_{k}(\q,i\wn)^2 = \frac{d}{2} \int_\q \Bigl(1+R'_k+\frac{2}{d}\q^2 R''_k \Bigr) G_{k}(\q,i\wn) - 2 v_d |\q|^d(1+R'_k) G_{k}(\q,i\wn) \bigl|_{|\q|=0}^{|\q|=\infty} , 
\eeq
which is obtained by an integration by part noting that $\partial_{\q^2}G_{k}=-(1+R'_k)G_{k}^2$, Eq.~(\ref{largeN4}) becomes  
\begin{align} 
K(i\wm) ={}& \frac{4}{d} T \sum_{\wn} \int_\q \q^2 \{ G(\q,i\wn) [G(\q,i\wn+i\wm) -G(\q,i\wn) ] \nonumber \\ &
- (1+R'_\Lamb)^2 G_{\Lamb}(\q,i\wn) [ G_{\Lamb}(\q,i\wn+i\wm) - G_{\Lamb}(\q,i\wn) ] \} ,
\label{largeN5} 
\end{align}
since the boundary term 
\beq
\frac{8 v_d}{d} \intw |\q|^d [ - G(\q,i\wn) + (1+R'_\Lamb) G_{\Lamb}(\q,i\wn) ] \bigl|_{|\q|=0}^{|\q|=\infty} 
= \frac{4 v_d}{d}  |\q|^d \biggl[ - \frac{1}{(\q^2+\Delta^2)^{1/2}} +  \frac{1+R'_\Lamb}{(\q^2+R_\Lamb+\Delta^2_\Lamb)^{1/2}} \biggr] \biggl|_{|\q|=0}^{|\q|=\infty}
\label{largeN5a} 
\eeq
vanishes (we have set $T=0$ in~(\ref{largeN5a})).The factor $v_d=1/[2^{d+1} \pi^{d/2}\Gamma(d/2)]$ originates from angular integrals. The contribution of the $k=0$ propagator in~(\ref{largeN5}) yields the known result\cite{Sachdev_book}
\beq
K(i\wm) = - \frac{\wm^2}{2} \int_\q \frac{\q^2}{(\q^2+\Delta^2)^{3/2} (\wm^2+4\q^2+4\Delta^2)} 
= - \frac{\wm^2}{24\pi\Delta} + \calO\left(\frac{\wm^4}{\Delta^3}\right) 
\label{largeNdisexpr} 
\eeq
for $T=0$ and $d=2$, i.e., $\Psi_B(i\wm=0,\rho=0)=-1/24\pi\Delta$. The contribution of the propagator $G_\Lamb$, of relative order $|\wm|/\Lamb$, can be ignored in the scaling limit. From~(\ref{Cdis}) we thus obtain 
\begin{equation}
\Cdis = \frac{\hbar\sig_q}{12\Delta} ,
\end{equation} 
restoring physical units.

At the QCP, $\Delta\to 0$. Still ignoring the $G_\Lambda$ terms, Eq.~(\ref{largeN5a}) yields the universal value $K(i\wm)=-|\wm|/16$, i.e.,\cite{Cha91}  
\beq
\sig(i\wm) = \frac{\pi}{8}\sig_q , 
\eeq
Note that at the QCP, $\Psi_B(i\wm,\rho=0)$ behaves as $1/|\wm|$ and $X_1(\rho=0)=-\Psi_B(i\wm=0,\rho=0)$ diverges.

\subsubsection{Ordered phase} 

In the ordered phase, the transverse and longitudinal propagators are given by\cite{Dupuis11}  
\begin{equation}
G_{k,\rm T}(\q,i\wn) = \frac{1}{\wn^2+\q^2+R_k} , \quad 
G_{k,\rm L}(\q,i\wn) = \frac{\frac{3N}{u_0} + \frac{N}{2} \Pi_k(\q,i\wn) }{(\wn^2+\q^2+R_k) \left( \frac{3N}{u_0} + \frac{N}{2} \Pi_k(\q,i\wn) \right) + 2\rho_{0,k} } , 
\end{equation}
where 
\begin{equation}
\Pi_k(\p,i\wm) = T\sum_{\wn} \int_\q G_{k,\rm T}(\q,i\wn) G_{k,\rm T}(\p+\q,i\wm+i\wn) .
\end{equation}

For a class B generator, the calculation of the conductivity is similar to the case of the disordered phase with $\Delta_k=0$; one finds Eq.~(\ref{largeN5}) with $G$ replaced by $G_{\rm T}$, which gives $K_{\rm B}(i\wm) = -|\wm|/16$ and  
\begin{equation}
\sigB(i\wm) = \frac{\pi}{8} \sigma_q .
\label{sigBlargeN}
\end{equation}
$\Psi_B(i\wm,\rho_0)$ behaves as $1/|\wm|$ and $X_1(\rho_0)$ is ill-defined in the whole ordered phase. 

For a class A generator, 
\begin{align}
K_{\rm A}(i\wm) ={}& - 2\rho_0 + T \sum_{\wn} \int_\q \biggl\{ \frac{4}{d} \q^2 [ G_{\rm L}(\q,i\wn) G_{\rm T}(\q,i\wn+i\wm) 
- (1+R'_\Lamb)^2 G_{\Lamb,\rm L}(\q,i\wn) G_{\Lamb,\rm T}(\q,i\wn+i\wm) ] \nonumber \\ & 
- G_{\rm L}(\q,i\wn) - G_{\rm T}(\q,i\wn) 
+ \Bigl(1+R'_\Lamb+\frac{2}{d}\q^2 R''_\Lamb \Bigr) [ G_{\Lamb,\rm L}(\q,i\wn) + G_{\Lamb,\rm T}(\q,i\wn) ]
\biggr\} .
\label{app_largeN_sig2} 
\end{align} 
Since $K_{\rm A}(i\wm)-K_{\rm B}(i\wm)= 2\rho_0[\wm^2\Psi_C(i\wm,\rho_0)-\Psi_A(i\wm,\rho_0)]$, we deduce 
\begin{multline}
2\rho_0 [ \Psi_A(i\wm,\rho_0) - \wm^2\Psi_C(i\wm,\rho_0) ] =  2\rho_0  
+ T \sum_{\wn} \int_\q \biggl\{ [ G_{\rm L}(\q,i\wn) - G_{\rm T}(\q,i\wn) ] 
\Bigl[ 1 - \frac{4}{d} \q^2 G_{\rm T}(\q,i\wn+i\wm) \Bigl] \\
- [ G_{\Lamb,\rm L}(\q,i\wn) - G_{\Lamb,\rm T}(\q,i\wn) ] 
\Bigl[ 1+R'_\Lamb+\frac{2}{d}\q^2 R''_\Lamb - \frac{4}{d} \q^2 (1+R'_\Lamb)^2 G_{\rm T}(\q,i\wn+i\wm) \Bigr] \biggr\} . 
\end{multline}
Thus, to leading order in the large-$N$ limit, $\Psi_A(i\wm,\rho_0)=1$ and $\Psi_C(i\wm,\rho_0)=0$. For $\wm\to 0$, $K_{\rm A}(i\wm)=\rho_s/\wm$ with $\rho_s=2\rho_0$, which yields  
\begin{equation}
\frac{\Cdis}{\sig_q^2 \Lord} = \frac{\pi}{6} \frac{\rho_s}{\Delta} = \frac{N}{24} ,
\label{CoverLlargeN}
\end{equation}
where the last result in~(\ref{CoverLlargeN}) is deduced using $\rho_s/\Delta=N/4\pi$.\cite{no5} It does not seem possible to obtain a simple expression for $\Psi_C$ to order $\calO(1/N)$. Nevertheless numerical evaluation of Eq.~(\ref{app_largeN_sig2}) shows that $K_A(i\wm)$ has no $\calO(\wm)$ term\footnote{This indicates that the $1/|\wm|$ divergence of $\Psi_B$ is cancelled by that of $2\rho_0\Psi_C$.} so that $\sig_A(\w)=\rho_s/(\w+i0^+)+\calO(\w)$, in agreement with perturbative results at finite $N$.\cite{Podolsky11}
\end{widetext}

\subsection{RG equations}  
\label{sec_largeN:subsec_nprg}

In the large-$N$ limit, the effective action $\Gamma$ is of order $N$, and the fields $\phibf$ and $\A$ of order $\sqrt{N}$ and one, respectively. Hence $Y_k$ is of order $1/N$ and can be neglected, $W_k$, $Z_k$ and $X_{2,k}$ are of order one, and $X_{1,k}$  of order $N$. 

To leading order,
\begin{equation} 
\dt Z_k = - \frac{N}{2} Z'_k \int_\q \dot R_k G^2_{k,\rm T} ,
\end{equation}
where the prime denotes a $\rho$ derivative and we omit to write the $\rho$ and $\q$ dependence of various functions ($\q$ is a $D$-dimensional vector). Since $Z_\Lambda=1$, $Z_k=1+\calO(1/N)$. The momentum dependence of the transverse propagator $G_{k,\rm T}=(\q^2+W_k+R_k)^{-1}$ is not renormalized in the large-$N$ limit. 

The equation for the potential then becomes 
\begin{equation}
\dt W_k = - \frac{N}{2} W'_k  \int_\q \dot R_k G^2_{k,\rm T} . 
\end{equation}
To solve this equation we set $W=W_k(\rho)$ and use the variables $(k,W)$ instead of $(k,\rho)$.\cite{Blaizot06,Rose15} Introducing the function $g_k(W)=\rho$ and using $g_k'(W)=1/W'_k(\rho)$, one obtains 
\begin{equation}
\dt g_k(W) = - \frac{1}{W'_k(\rho)} \dt W_k(\rho) = \frac{N}{2} \int_\q \dot R_k G_{k,\rm T}^2 . 
\label{dtgk} 
\end{equation}
Since $k$ and $W$ are independent variables, this equation can be rewritten as 
\begin{equation}
\dt g_k(W) = - \frac{N}{2} \dt \int_\q \frac{1}{\q^2+W+R_k} , 
\end{equation}
where both sides are total derivatives, and we obtain 
\begin{multline}
g_k(W)-g_\Lambda(W) \\ 
= - \frac{N}{2} \int_\q \biggl( \frac{1}{\q^2+W+R_k} - \frac{1}{\q^2+W+R_\Lambda} \biggr) . 
\end{multline}
For $R_\Lambda(\q)\to\infty$ and with $g_\Lamb(W)=(3N/u_0)(W-r_0)$,  
\begin{equation}
W_k(\rho) = r_0 + \frac{u_0\rho}{3N} + \frac{u_0}{6} \int_\q  \frac{1}{\q^2+W_k(\rho)+R_k} , 
\label{WlargeN} 
\end{equation}
which reproduces the known result in the large-$N$ limit for $k\to 0$.\cite{Zinn_book} 

To leading order the flow of $X_{1,k}$ reads 
\begin{equation}
\dt X_{1,k} = - \frac{N}{2} X'_{1,k}  \int_\q \dot R_k G^2_{k,\rm T} . 
\end{equation}
Since $X_{1,\Lamb}=0$, $X_{1,k}$ is in fact of order one. Thus, to leading order, 
\begin{equation}
\dt X_{1,k} = - \frac{N}{2} X'_{1,k}  \int_\q \dot R_k G^2_{k,\rm T} + \tdt I_{X_1} , 
\label{dX1}
\end{equation}
where $I_{X_1}$ collects all terms of order one. In Appendix~\ref{app_largeN_IX1} we show that 
\begin{equation}
I_{X_1} = \int_\q \frac{G_{k,\rm T} (\partial^2_{\q^2} G_{k,\rm T}) - (\partial_{\q^2} G_{k,\rm T}){}^2}{3 G_{k,\rm T}^2}.
\label{IX1} 
\end{equation}
With $(k,W)$ as variables we find 
\begin{equation}
\dt X_{1,k}\bigl|_W = \dt X_{1,k}\bigl|_\rho + X'_{1,k} \dt g_k = \dt I_{X_1} , 
\end{equation}
using~(\ref{dtgk}) and $\tdt I_{X_1}=\dt I_{X_1}$ since $W$ is $k$ independent. We then obtain 
\begin{align}
X_{1,k} ={}& \int_\q \biggl( \frac{G_{k,\rm T} (\partial^2_{\q^2} G_{k,\rm T}) - (\partial_{\q^2} G_{k,\rm T}){}^2}{3 G_{k,\rm T}^2} \nonumber \\
&{}- \frac{G_{\Lambda,\rm T} (\partial^2_{\q^2} G_{\Lambda,\rm T}) - (\partial_{\q^2} G_{\Lambda,\rm T}){}^2}{3 G_{\Lambda,\rm T}^2} \biggl) . 
\end{align}
For $k=0$ and $R_\Lamb\to\infty$, this gives 
\begin{equation}
X_{1,{k=0}}(\rho) = \third \int_\q G^2_{k=0,\rm T} = \frac{1}{24\pi \sqrt{W_{k=0}(\rho)}} ,
\label{X1largeN} 
\end{equation} 
using $[G_{\rm T}(\partial^2_{\q^2} G_{\rm T})-(\partial_{\q^2} G_{\rm T})^2]/3G_{\rm T}^2=G_{\rm T}^2/3$ for $G_{\rm T}\equiv G_{k=0,\rm T}=(\q^2+W)^{-1}$.
In the disordered phase, where $W_{k=0}(0)$ is equal to the excitation gap $\Delta$, Eq.~(\ref{X1largeN}) yields $\Psi_B(i\wm=0,\rho=0)=-1/24\pi\Delta$ in agreement with the direct calculation of Sec.~\ref{sec_largeN:subsec_direct}. In the ordered phase, $X_{1,k=0}$ diverges since $W_k(\rho_{0,k})=0$. This divergence holds for all values of $N$ and reflects the divergence of $\Psi_B(\p,\rho_0)$ in the limit $\p\to 0$. This prevents us to compute $\sigB(\w)$ from the derivative expansion. We shall come back to this issue in Sec.~\ref{sec_num}.  

To leading order the flow of $X_{2,k}$ reads 
\begin{equation}
\dt X_{2,k} =  - \frac{N}{2} X'_{2,k}  \int_\q \dot R_k G^2_{k,\rm T} . 
\end{equation}
Since $X_{2,\Lamb}=0$, $X_{2,k}$ is in fact of order $1/N$. Unlike the case of $X_{1,k}$, no simple expression is obtained for $X_{2,k}$.

\section{Numerical solution of the flow equations} 
\label{sec_num} 

The QCP manifests itself as a fixed point of the RG equations provided we use dimensionless equations where all quantities are expressed in units of the running scale $k$. We therefore introduce the following dimensionless variables (with $D=3$) 
\begin{equation}
\begin{gathered} 
\tilde p=k^{-1} p, \quad \trho = v_D^{-1} Z_k k^{2-D} \rho , \\
\tilde A_\mu = k^{-1} A_\mu, \quad \tilde F_{\mu\nu} = k^{-2} F_{\mu\nu}  
\end{gathered}
\end{equation}
and functions
\begin{equation}
\begin{gathered} 
\tilde{U}_k(\trho) =v_D^{-1}k^{-D}U_k(\rho), \quad 
\tilde{W}_k(\trho) =Z_k^{-1}k^{-2}W_k(\rho),\\
\tilde{Z}_k(\trho) =Z_k^{-1}Z_k(\rho),\quad 
\tilde{Y}_k(\trho) =v_D Z_k^{-2} k^{D-2} Y_k(\rho),\\
\tilde{X}_{1,k}(\trho) =v_D^{-1}k^{4-D} X_{1,k}(\rho),\quad 
\tilde{X}_{2,k}(\trho) =Z_k^{-1} k^{2} X_{2,k}(\rho) ,
\end{gathered}
\label{vdim}
\end{equation}  
where $v_D$ is defined after~(\ref{largeN5a}). 

The flow equations are solved numerically using the explicit Euler method. It is convenient to introduce the (negative) RG ``time'' $t=\ln(k/\Lambda)$. $\Lamb=1$ and $u_0$ is typically equal to $100 v_D$. Both $t$ and $\trho$ variables are discretized with finite-size steps $\Delta t = -5\times 10^{-4}$ and $\Delta \trho = 0.1$, respectively. We use a fixed $\trho$ grid with $0\leqslant \trho \leqslant \trho_{\text{max}}=8N$. Momentum integrals are computed using Simpson's rule with 60 points and an upper cutoff $\tilde q_{max}=6$; this cutoff is justified by the presence of the regulator $R_k$. Stability of the results with respect to the various parameters (step size, $\trho_{\text{max}}$, numerical accuracy of the integrals) has been verified.

As for the regulator, we take the exponential cutoff function 
\begin{equation}
R_k(\q^2) = Z_k \q^2 r\left(\frac{\q^2}{k^2}\right) , \quad r(y) = \frac{\alpha}{e^y-1} , 
\label{rdef} 
\end{equation}
where $\alpha$ is a constant of order one and $Z_k\equiv Z_k(\rho_{0,k})$. In the derivative expansion, as in any approximation scheme, the results slightly depend on the shape of the regulator. We minimize this dependence by using the ``principle of minimal sensitivity'', that is, we search for a local extremum of the physical quantities taken as functions of the regulator. In practice, for each physical quantity ${\cal Q}$, this amounts to finding the optimal value $\alpha_{\rm opt}$ such that $d{\cal Q}/d\alpha|_{\alpha_{\rm opt}}=0$.\footnote{$\alpha_{\rm opt}$ is close to $2.25$ for the critical exponents and the ratio $\rho_s/\Delta$.}

In the ordered phase, two issues arise. First, since the location of the minimum of the potential converges towards a finite value $\rho_0=\lim_{k\to 0}\rho_{0,k}$, its dimensionless counterpart diverges as $\trho_{0,k}\sim 1/Z_k k$ when $k\to 0$ (here $D=3$). Furthermore, since $W_k(\rho)$ is a monotonously increasing function and $W_k(\rho_{0,k})=0$,  $W_k(\rho)$ is negative for $\rho<\rho_{0,k}$ and there is no guarantee that the transverse propagator $G_{k,T}(\p,\rho=0)=[Z_k(0)\p^2+ W_{k}(0)+R_k(\p)]^{-1}$ remains positive definite. To second order of the derivative expansion, the propagator indeed exhibits a pole for a certain value $k_c$ of the running momentum scale, which prevents to continue the flow for $k<k_c$. 

To circumvent these difficulties, we proceed as follows. When $\trho_{0,k}$ comes too close to $\trho_{\rm max}$ (e.g., $\trho_{0,k}>0.75\trho_{\rm max}$), we follow the minimum of the potential while keeping dimensionless variables. To that effect, we perform the $k$-dependent change of variables $\trho\to\delta\trho=\trho-\trho_{0,k}$ and switch to a grid $[\delta\trho_\text{min},\delta\trho_\text{max}]$ with $\delta\trho_\text{min}<0<\delta\trho_\text{max}$ such that the minimum of the potential $\delta \trho=0$ is within the grid. In the original dimensionless variables, this corresponds to a $\trho$ window of fixed width that follows $\trho_{0,k}$ as it flows to infinity. In the dimensionful variables, the grid is centered about $\rho_{0,k}$ with a $k$-dependent width $v_dZ_k^{-1}k^{d-2}(\delta\trho_\text{max}-\delta \trho_\text{min})$ which goes to zero with $k$. Our approach is therefore reminiscent of the usual field expansion about $\rho_{0,k}$,\cite{Berges02,Delamotte12} the main difference being that we retain the full $\trho$ dependence of the functions. In the new grid, no instability arises; the flow can be continued to arbitrary small values of $k$ and all dimensionless functions (e.g. $\tilde W_k(\delta\trho)$) reach the ordered-phase fixed point. 

After the change of variables, $\trho_{0,k}$ intervenes explicitly in the flow equations. While in the original grid $\trho_{0,k}$ is determined by the condition $\tilde W_k(\trho_{0,k})=0$, in the new grid one has to keep track of the flow of $\trho_{0,k}$ separately. This is obtained by differentiating with respect to time the identity $\tilde W_k(\delta \trho = 0)\equiv\tilde W_k(\trho_{0,k})=0$. 

The equations are solved for several sets of initial conditions $(r_0,u_0)$. For a given value of $u_0$, the QCP can be reached by fine tuning $r_0$ to its critical value $r_{0c}$. The universal regime near the QCP can then be studied by tuning $r_0$ slightly away from $r_{0c}$. Universality of various quantities can be checked by changing the value of $u_0$. 

\begin{figure} 
\centerline{\includegraphics{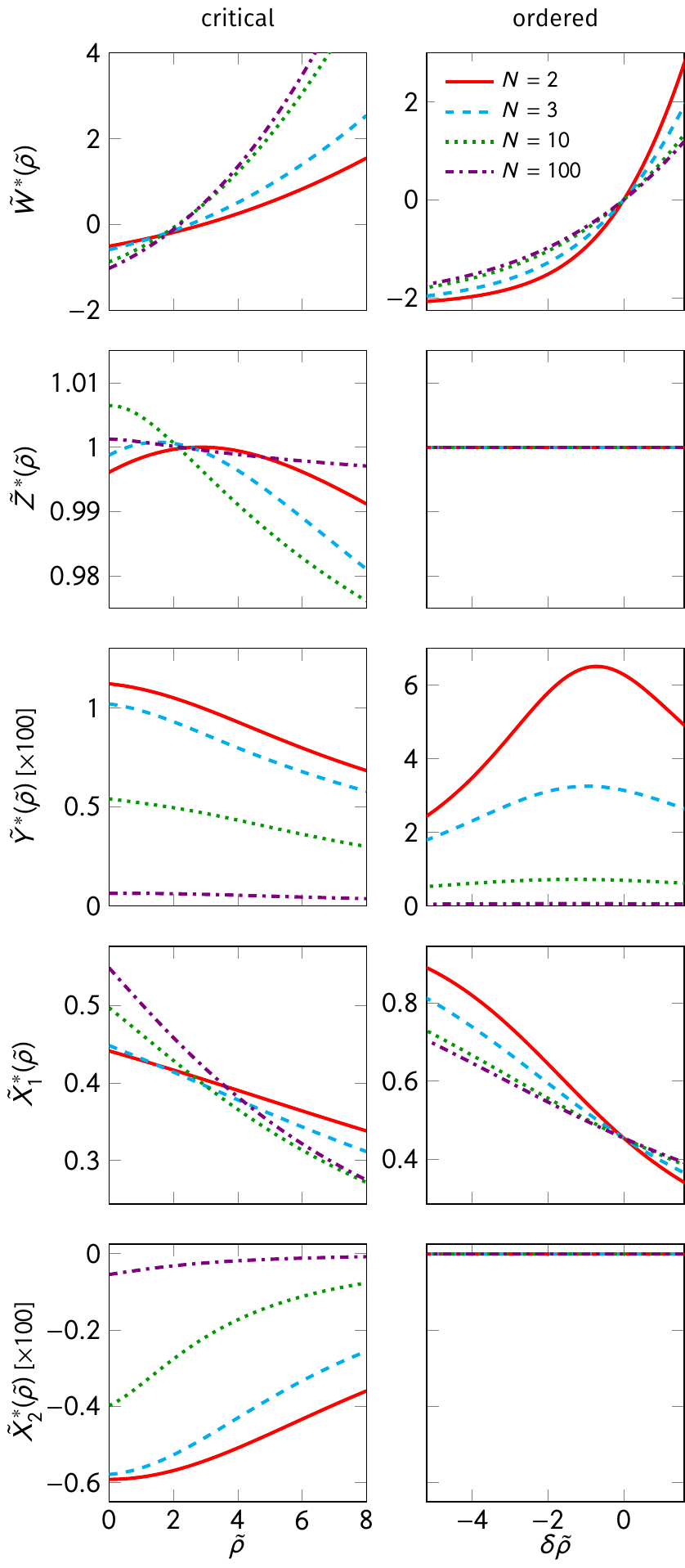}}
\caption{Dimensionless fixed-point solutions of the flow equations at the QCP (left) and the ordered phase (right) for several values of $N$. The quantities are respectively expressed as a function of $\trho$ and $\delta\trho=\trho-\trho_{0,k}$ at the critical point and in the ordered phase. In the latter case, the fixed-point solutions $\tilde Z^*(\delta\trho)=1$ and $\tilde{X}^*_2(\delta\trho)=0$ are independent of $N$.}
\label{fig_FPpot}
\end{figure}

The fixed-point values of the $\trho$-dependent functions $\tilde W$, $\tilde Z$ and $\tilde Y$  are shown in Fig.~\ref{fig_FPpot}. At criticality we reproduce known results of the derivation expansion to second order.\cite{Gersdorff01,Jakubczyk14} To our knowledge the results in the ordered phase, showing the fixed-point functions of the $\delta\trho$ variable, are new. Fig.~\ref{fig_FPpot} also shows the functions $\tilde X_1$ and $\tilde X_2$ that determine the conductivity.

\subsection{Critical exponents and ratio $\rho_s/\Delta$} 

Let us first verify that we reproduce known values of the critical exponents $\nu$ and $\eta$. The anomalous dimension $\eta$ is directly obtained from $\lim_{k\to 0} \eta_k$ when $r_0=r_{0c}$, where $\eta_k=-\dt\ln Z_k$. Since $\dt Z_k$ can be obtained from $\dt\Gamma^{(2,0)}_k$, $\dt\Gamma^{(1,1)}_k$ or $\dt\Gamma^{(0,2)}_k$, we obtain three different equations for $\eta_k$. They all give the same result within numerical accuracy. The correlation-length exponent $\nu$ is deduced from the behavior of $\trho_{0,k}\simeq \trho^*_0+Ce^{-t/\nu}$ for long times $|t|$ such that the flow leaves the critical regime and enters either phase (since the condition $r_0=r_{0c}$ is never exactly fulfilled the RG trajectories will always eventually flow away from the fixed point with an escape rate given by $1/\nu$). Our results agree with previous NPRG calculations\cite{Benitez12,Rose15} in the BMW approximation and are in reasonable agreement with Monte Carlo estimates (see Tables~\ref{table_nu} and \ref{table_eta}).

\begin{table}
\centering
\setlength{\tabcolsep}{8pt}
\caption{Critical exponent $\nu$  
obtained in the NPRG approach with the derivative expansion (NPRG DE, this work) or the BMW approximation (NPRG BMW), compared to Monte Carlo (MC) simulations.}
\begin{tabular}{cccc}
\hline \hline
$N$ & NPRG DE & NPRG BMW\cite{Rose15} & MC \\
\hline
2 & 0.668 & 0.673 & 0.6717(1)~[\onlinecite{Campostrini06}] \\
3 & 0.706 & 0.714 & 0.7112(5)~[\onlinecite{Campostrini02}]\\
4 & 0.741 & 0.754 & 0.749(2)~[\onlinecite{Hasenbusch01}]\\
5 & 0.774 & 0.787 & \\
6 & 0.803 & 0.816 & \\
8 & 0.848 & 0.860 & \\
10 & 0.879 & 0.893 & \\
100 & 0.989 & 0.990 & \\
1000 & 0.999 & 0.999 & \\
\hline \hline
\end{tabular}
\label{table_nu}
\caption{Same as Table~\ref{table_nu} but for the anomalous dimension $\eta$ (defined by $\eta=\lim_{k\to 0}\eta_k$ when $r_0=r_{0c}$).}
\begin{tabular}{cccc}
\hline \hline
$N$ & NPRG DE & NPRG BMW\cite{Rose15} & MC\\
\hline
2 & 0.0467 & 0.0423 & 0.0381(2)~[\onlinecite{Campostrini06}] \\
3 & 0.0463 & 0.0411 & 0.0375(5)~[\onlinecite{Campostrini02}] \\
4 & 0.0443 & 0.0386 & 0.0365(10)~[\onlinecite{Hasenbusch01}] \\
5 & 0.0413 & 0.0354 &    \\
6 & 0.0381 & 0.0321 &    \\
8 & 0.0319 & 0.0264 &    \\
10 & 0.0270 & 0.0220 &    \\
100 & 0.00296 & 0.00233 &    \\
1000 & 0.000296 & 0.000233 &    \\
\hline \hline
\end{tabular}
\label{table_eta}
\setlength{\tabcolsep}{2pt}
\caption{Universal ratio $\rho_s/(N\Delta)$ obtained from the NPRG in the derivation expansion (NPRG DE, this work). Also shown are the previous NPRG results obtained from the BMW approximation\cite{Rose15} (NPRG BMW) and from Monte Carlo simulations\cite{Gazit13a} (MC). The exact result in the limit $N\to\infty$ is $1/4\pi\simeq 0.0796$.}
\begin{tabular}{ccccccccc}
\hline \hline
$N$ & 1000 &  10 & 8 & 6 & 4 & 3 & 2 \\
\hline
NPRG DE & 0.0795 & 0.0810  & 0.0846 & 0.0935  & 0.118  & 0.147  & 0.207  \\
NPRG BMW\cite{Rose15}  & 0.0796 &  0.0803 & 0.0829  & 0.0903  & 0.111 & 0.137 & 0.193 \\
MC\cite{Gazit13a} & & &   &  & & 0.114 & 0.220  \\
\hline \hline
\end{tabular}
\label{table_rhos}
\end{table}

The universal ratio $\rho_s/\Delta$ is shown in table~\ref{table_rhos}. Here $\Delta$ denotes the excitation gap in the disordered phase at the point located symmetry wrt the point of the ordered phase where $\rho_s$ is computed. Our results are in good agreement with previous results obtained in the BMW approximation.\cite{Rose15}

\subsection{Conductivity} 
\label{sec_num:subsec_cond} 

As pointed out in Sec.~\ref{sec_nprg:subsec_de}, the derivative expansion allows one to calculate correlation functions only in the limit $|\p|\ll k$. Any correlation function at nonzero momentum $\p$ seems therefore to be out of reach since it requires to take the limit $k\to 0$ and therefore consider the case $k\ll|\p|$ where the derivative expansion is not valid. It is however well known that the derivative expansion is often valid beyond its apparent domain of validity and allows one to compute correlation functions at finite momentum. Consider for example the transverse two-point vertex $\Gamma^{(2,0)}_{k,\rm T}(\p,\rho_{0,k})=Z_k \p^2+W_k(\rho_{0,k})$ evaluated at the minimum of the effective potential. In the disordered phase ($\rho_{0,k}=0$), the flow of $Z_k$ and $W_k(0)$ stops when $k$ becomes smaller than the inverse of the correlation length $\xi=(Z_{k=0}/W_{k=0}(0))^{1/2}$. The excitation gap $\Delta\equiv\xi^{-1}$ thus acts as an infrared regulator and the derivative expansion is valid for all momenta $|\p|\ll\Delta$ regardless of the value of $k$. In the ordered phase, where  $W_k(\rho_{0,k})=0$, the flow of $\rho_{0,k}$ and $Z_k$ stops when $k$ becomes smaller than the inverse of the Josephson length $\xi_J$,\cite{Josephson66,no6} with the stiffness defined by $\rho_s=\lim_{k\to 0}2Z_k\rho_{0,k}$. Hence $\lim_{k\to 0}\Gamma^{(2,0)}_{k,\rm T}(\p,\rho_{0,k})=\rho_s\p^2/2\rho_0$, which turns out to be the exact result for the transverse vertex to order $\p^2$. Thus the derivative expansion is valid for $|\p|\ll \xi_J^{-1}$ regardless of the value of $k$. Quite generally, we expect the derivative expansion to remain valid for small, but nonzero, momentum whenever the $k$-dependent functions relevant for the computation of the correlation function reach a finite limit when $k\to 0$.\footnote{An irrelevant function vanishing as a power of $k$ would also make the results of the derivative expansion invalid in the regime $k\ll|\p|$.} At criticality, where $\rho_{0,k}\to 0$  and $Z_k(0)\sim k^{-\eta}$, the derivative expansion yields $\Gamma^{(2,0)}_{k,\rm T}(\p,0)\sim k^{-\eta}\p^2$ and is valid only for $|\p|\ll k$. It is nevertheless possible to retrieve the momentum dependence of the vertex in the limit $k\ll|\p|$ by setting $k\sim|\p|$ in the derivative expansion result, noting that $\p$ acts as an effective infrared cutoff, which reproduces the known result $\Gamma^{(2,0)}_{k,\rm T}(\p,0)\sim |\p|^{2-\eta}$ for the two-point vertex at criticality (note however that the overall factor cannot be obtained by this simple trick). We are now in a position to discuss the derivative expansion results for the conductivity.

\subsubsection{Universal ratio $\Cdis/\Lord\sig_q^2$} 

\begin{table}
\centering
\setlength{\tabcolsep}{8pt}
\caption{Ratio $\hbar\sig_q/2\pi \Cdis\Delta$ and $\Cdis/N\Lord\sig_q^2$ obtained from the NPRG approach. The exact results for $N\to\infty$ are  $6/\pi\simeq 1.90986$ and $1/24\simeq 0.41667$, respectively.}
\begin{tabular}{ccc}
\hline \hline
$N$ & $\hbar\sig_q/2\pi \Cdis\Delta$ & $\Cdis/N\Lord\sig_q^2$ \\
\hline
2 & 1.98 & 0.105\\
3 &1.98& 0.0742\\
4 &1.98& 0.0598\\ 
5 &1.97& 0.0520\\ 
6 &1.97& 0.0475\\ 
8 &1.96&  0.0731\\
10 &1.96&  0.0415\\
100 &1.92&  0.0413\\
1000 & 1.91 & 0.0416\\
\hline \hline
\end{tabular}
\label{table_CoverL}
\end{table}

Let us start with the conductivities $\sig(\w)$ in the disordered phase and $\sigA(\w)$ in the ordered phase, in the limit $\w\to 0$. The former is determined by $X_{1,k}(\p=0,\rho=0)$, the latter by $\rho_{s,k}$. Both quantities reach a finite limit when $k\to 0$ so that we are in the case where the derivative expansion remains reliable for $k\to 0$ (and $\w$ nonzero). 
In table~\ref{table_CoverL} we show the universal ratio $\hbar\sig_q/2\pi \Cdis\Delta$ and $\Cdis/N\Lord\sig_q^2$. In the limit $N\to\infty$ our results agree with the exact results $6/\pi$ and $1/24$. For $N=2$, we find $\hbar\sig_q/2\pi \Cdis\Delta\simeq 1.98$, in reasonable agreement with the Monte Carlo result 2.1(1).\cite{Gazit14}

\subsubsection{Universal conductivity at QCP} 
\label{sec_num:subsubsec_condQCP}

At the QCP, we expect the zero-frequency conductivity $\sig^*$ to be given (in units of $\sig_q$) by a universal number.\cite{Fisher90} In the disordered phase, the $k$-dependent conductivity is given by  
\begin{equation}
\sig_k(i\wn) = 2\pi\hbar \sig_q \wn X_{1,k}(\rho_{0,k}) , 
\label{sigstar}
\end{equation}
where $X_{1,k}(\rho)=v_3\tilde X_{1,k}(\trho)/k$. Since both $\trho_{0,k}$ and $\tilde X_{1,k}(\trho)$ reach fixed-point values at the QCP (Fig.~\ref{fig_FPpot}), $X_{1,k}(\rho=0)\sim 1/k$ diverges in the limit $k\to 0$. As explained at the beginning of Sec.~\ref{sec_num:subsec_cond} the derivative expansion does not allow us to obtain the conductivity at nonzero frequency in such a case (the limit $k\to 0$ at fixed $\wn$ in~(\ref{sigstar}) is divergent). Setting $k\sim |\wn|$ in Eq.~(\ref{sigstar}), we see that $\sig_k(i\wn)$ reaches a $\wn$-independent limit $\sig^*$ if we approximate $\tilde X_{1,k}(\trho_{0,k})$ by its fixed-point value $\tilde X^{*\rm cr}_{1}(\trho^*_0)$, but the precise value of $\sig^*$ cannot be obtained. 

In the derivative expansion, the function $\Psi_{B,k}(\p,\rho)$ is approximated by $-X_{1,k}(\rho)$ [Eqs.~(\ref{vert2})]. The $1/k$ divergence of $X_{1,k}(\rho=0)$ suggests that at the QCP $\Psi_{B,k=0}(\p,\rho=0)$ is in fact not analytic in $\p^2$ but behaves as $1/|\p|$ for $\p\to 0$, as can be shown explicitly in the large-$N$ limit (Sec.~\ref{sec_largeN}). (Note however that $\Psi_{B,k}(\p,\rho)$ is an analytic function of $\p$ for $|\p|\ll k$.)

\subsubsection{Universal conductivity $\sigB$} 

\begin{figure} 
\centerline{\includegraphics[width=7.5cm]{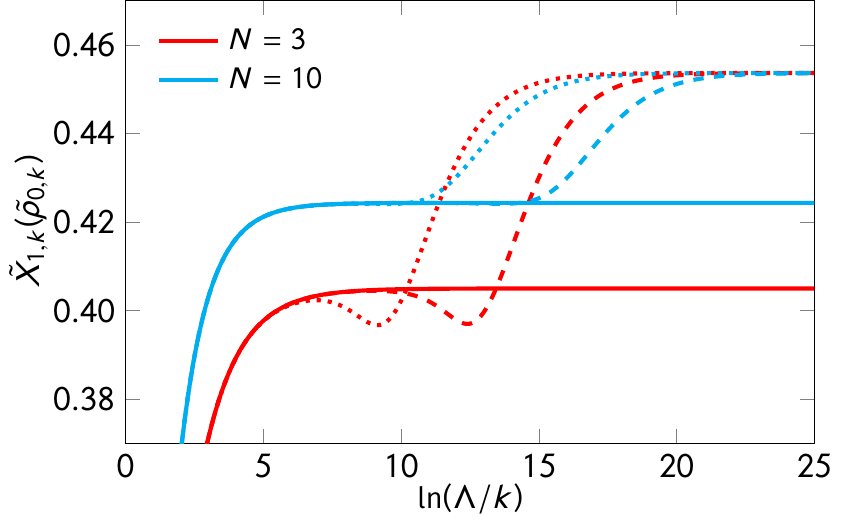}}
\caption{Flow of $\tilde X_{1,k}(\trho_{0,k})$ in the ordered phase for $N=3$ and $N=10$, and three different initial conditions, with $\alpha=2.25$. The solid lines show the system at criticality where, after a transient regime, $\tilde X_{1,k}(\trho_{0,k})$ becomes very close to its fixed point value $\tilde X_{1,k}^{*\rm cr}(\trho^*_{0})$. The dashed and dotted lines correspond to the ordered phase: $\tilde X_{1,k}(\trho_{0,k})$ is first attracted by the critical fixed point; for $k\sim\xi_J^{-1}$ there is a crossover to a regime dominated by the Goldstone modes and $\tilde X_{1,k}(\trho_{0,k})\simeq \tilde X_{1,k}^{*\rm ord}(\trho_{0,k})$.}
\label{fig_sigB} 
\end{figure}

The conductivity tensor in the ordered phase is defined by $\sigA(\w)$ and $\sigB(\w)$ (Sec.~\ref{subsec_eaf:subsubsec_condQM}). $\sigA(\w\to 0)$ is entirely determined by the stiffness $\rho_s$. $\sigB(\w\to 0)$ is obtained from $X_{1,k}(\rho_{0,k})$ and its expression is the same as that of the conductivity~(\ref{sigstar}) at the QCP. Since $\tilde X_{1,k}(\delta\trho)$ reaches a fixed-point value $\tilde X^{*\rm ord}_{1}(\delta\trho)$ in the ordered phase (Fig.~\ref{fig_FPpot}), $X_{1,k}(\rho_{0,k})\sim 1/k$ diverges in the limit $k\to 0$ and we come to the conclusion that $\Psi_{B,k=0}(\p,\rho_0)$ behaves as $1/|\p|$ for $\p\to 0$ (a property that can be explicitly verified in the large-$N$ limit). Following the same reasoning as in Sec.~\ref{sec_num:subsubsec_condQCP} we can nevertheless infer that $\sigB(\w\to 0)$ takes a universal value $\sigB^*$ independent of the distance $r_{0c}-r_0$ to the QCP. Furthermore, we find numerically that the fixed-point value $\tilde X^{*\rm ord}_{1}(\delta\trho=0)$ is independent of $N$, unlike the fixed-point value $\tilde X^{*\rm cr}_{1}(\trho^*_0)$ at the QCP.  

These results are illustrated in Figs.~\ref{fig_FPpot} and \ref{fig_sigB}. Since $\sigB$ is known in the large-$N$ limit [Eq.~(\ref{sigBlargeN})], this leads us to conjecture that $\sigB(\w\to 0)$ takes the $N$-independent universal value  
\begin{equation}
\sigB^* = \frac{\pi}{8} \sig_q .
\end{equation}
In the limit $N\to\infty$, $\sigB^*$ and $\sig^*$ are thus equal (Sec.~\ref{sec_largeN}) but differ for $N<\infty$ since $\sig^*$ is $N$ dependent while $\sigB$ is not. In the limit $N\to\infty$ we indeed find that $\tilde{X}_1^{*\rm ord}(\delta\trho=0)=\tilde{X}_1^{*\rm cr}(\trho^*_0)$. This follows not only from the equality $\sigB^*=\sig^*$ but also from the fact that the critical and ordered phase fixed points are entirely determined by the Goldstone modes and therefore similar when $N\to\infty$. For $N<\infty$, it is not possible to strictly identify $\sigB^*/\sig^*$ to $\tilde{X}_1^{*\rm ord}(\delta\trho=0)/\tilde{X}_1^{*\rm cr}(\trho^*_0)$. The latter ratio is shown in Table~\ref{table_sigBsigstar}.

\begin{table}
\centering
\setlength{\tabcolsep}{8pt}
\caption{Ratio $\tilde{X}_1^{*\rm ord}(\delta\trho=0)/\tilde{X}_1^{*\rm cr}(\trho^*_0)$ obtained from the NPRG approach for $\alpha=2.25$. (Note that $\sigma_B$ is not defined for $N=2$.) The exact result for $N\to\infty$ is $1$.} 
\begin{tabular}{rc}
\hline \hline
$N$ & $\tilde{X}_1^{*\rm ord}(\delta\trho=0)/\tilde{X}_1^{*\rm cr}(\trho^*_0)$ \\
\hline
2 & 1.1210\\
3 & 1.1201\\
4 & 1.1146\\ 
5 & 1.1068\\ 
6 & 1.0982\\ 
8 &  1.0821\\
10 &  1.0691\\
100 &  1.00739\\
1000 & 1.00073\\
\hline \hline
\end{tabular}
\label{table_sigBsigstar}
\end{table}

\subsubsection{Conductivity $\sigA$} 

Perturbative results at finite $N$ as well as the large-$N$ limit (Sec.~\ref{sec_largeN}) show that $\sigA(\w)=\rho_s/(\w+i0^+)+\calO(\w)$ in the ordered phase. The absence of $\calO(\w^0)$ term implies that the $1/|\p|$ divergence in $\Psi_B(\p,\rho)$ is canceled by a similar divergence in $2\rho_0\Psi_C(\p,\rho)$. To recover this result from the NPRG approach, we need $\tilde X_1^*(\trho_{0,k})+2\trho_{0,k}\tilde X_2^*(\trho_{0,k})=0$.\footnote{In the ordered phase and for $d=2$, $\trho_{0,k}\sim 1/k$ and $\tilde X_{2,k}(\trho_{0,k})\sim k$, so that $\trho_{0,k}\tilde X_{2,k}(\trho_{0,k})$ reaches a finite limit when $k\to 0$.}  
Although $\tilde X_1^*(\trho_{0,k})$ and $2\trho_{0,k}\tilde X_2^*(\trho_{0,k})$ have opposite sign, their sum does not vanish and is quite sensitive to the choice of the parameter $\alpha$. A proper calculation of $\sigA(\w)$ to order $\w$ requires to go beyond the derivative expansion. 

\section{Conclusion} 

The NPRG is well known as an efficient tool to calculate both universal and nonuniversal (e.g. in lattice models) quantities of quantum critical systems.\cite{Wetterich08,Rancon11a,Rancon11b,Rancon12a,Rancon12d,Rancon13a,Rancon13b,Rancon14,Rose15,Rancon16} In this paper we have shown, in the framework of the quantum O($N$) model, that the NPRG approach can be used to study transport properties. Our results are obtained from a derivative expansion to second order of the effective action in the presence of an external non-Abelian gauge field. 

We have calculated the universal ratio $\Cdis/\Lord\sig_q^2$ as a function of $N$, where $\Cdis$ is the capacitance of the (insulating) disordered phase and $\Lord$ the inductance of the (``superfluid'') ordered phase associated with the element $\sigA(\w)$ of the conductivity tensor. Our results are in agreement with a recent QMC calculation in the case $N=2$.\cite{Gazit14} 

Although the derivative expansion used in our work does not allow us to calculate the conductivity $\sig(\w\to 0)\equiv\sig^*$ at the QCP and the element $\sigB(\w\to 0)\equiv\sigB^*$ of the conductivity tensor in the ordered phase (see the discussion in Sec.~\ref{sec_num:subsec_cond}), our results strongly suggest that both $\sig^*$ and $\sigB^*$ are universal. While the existence of a universal value for $\sig^*/\sig_q$ is well known,\cite{Fisher90} to our knowledge the universal character of $\sigB^*/\sig_q$ has not been pointed out before. Our RG analysis leads us to conjecture that $\sigB^*=\frac{\pi}{8}\sig_q$ takes a $N$-independent universal value in the whole ordered phase. 

In a forthcoming paper we will show how an improved derivative-expansion scheme can be used to obtain the full frequency dependence of correlation functions, thus allowing us to compute $\sig^*$ and $\sigB^*$ and check whether our conjecture regarding the universal value of $\sigB^*$ is correct. 

Analytical continuation of imaginary-time data to real time with Pad\'e approximants works well at zero temperature\cite{Rose15,Rose16a} but not at finite temperature in the low-frequency regime $|\w|\lesssim T$. It has recently been pointed out that the NPRG enables, within some approximation schemes, to perform exactly both Matsubara-frequency sums and analytical continuation to real frequencies.\cite{Kamikado14,Tripolt14,Tripolt14a,Wambach14} Other methods to circumvent the difficulties in performing analytical continuation at finite temperature from numerical data have been proposed.\cite{Pawlowski15} Whether or not these proposals can be successfully implemented for the quantum O($N$) model is likely to determine whether the NPRG is a reliable method to compute finite-temperature transport properties.

\begin{acknowledgments}
We thank A. Tresmontant for useful discussion. 
\end{acknowledgments} 

\appendix

\section{Generators of the SO($N$) group}
\label{app_SON}

Any rotation matrix $O\in\mbox{SO}(N)$ can be written as 
\begin{equation}
O = \exp(\theta^a T^a) 
\end{equation}
where $\{\theta^a\}$ are real parameters and $\{T^a\}$ $N(N-1)/2$ generators of the SO($N$) group. The latter form a basis of the special orthogonal Lie algebra so($N$) defined as the set of skew-symmetric $N\times N$ matrices. They satisfy the commutation relations 
\begin{equation}
[T^a,T^b] = f_{abc} T^c ,
\end{equation}
where the structure constants $f_{abc}$ are real and antisymmetric under permutation of indices: $f_{abc}=-f_{bac}=-f_{cba}$, etc. For an infinitesimal rotation ($\theta^a \to 0$), 
\begin{equation}
O \simeq 1 + \theta^a T^a , 
\end{equation}
where $1$ in the rhs stands for the $N\times N$ identity matrix. 

A convenient basis of so($N$) is 
\begin{equation}
T^{IJ}_{ij} = - \delta_{I,i}\delta_{J,j} + \delta_{I,j}\delta_{J,i} ,
\label{eq:basisTIJ}
\end{equation}
where the integers $I$ and $J$ satisfy $1\leq I<J\leq N$. For $N=2$, there is a single generator defined by $T_{ij}=-\delta_{i,1}\delta_{j,2}+\delta_{i,2}\delta_{j,1}$, i.e., $T_{ij}=-\eps_{ij}$ with $\eps_{ij}$ the fully antisymmetric tensor, and any rotation matrix can be written as 
\begin{equation}
O = \exp(\theta T) = \left( 
\begin{matrix} 
\cos\theta & - \sin\theta \\ 
\sin\theta & \cos\theta 
\end{matrix}
\right) .
\end{equation}
For $N=3$, there are three generators: $T^{12}=-iS^3$, $T^{13}=iS^2$ and $T^{23}=-iS^1$, where the $(S^1,S^2,S^3)$ are spin-one matrices: $\sum_i (S^i)^2=2$ and $[S^j,S^k]=i\eps_{jkl}S^l$.

\section{Conductivity in the effective action formalism} 
\label{app_K} 

In this Appendix we generalize the definition of the response function $\Kmunuab$ to an arbitrary uniform field $\phibf$ and further discuss the $f$-sum rule~(\ref{fsum}). 

\subsection{Conductivity from 1PI vertices} 

In the effective action formalism the conductivity is defined by~(\ref{K1}).
Using~(\ref{phibardef}), one finds 
\begin{multline}
\frac{\bdelta^2 \Gamma[\phibf[\A],\A]}{\bdelta \Amua(\y) \bdelta \Anub(\y')} = \Gamzdmunuab[\y,\y';\phibf[\A],\A] \\
+ \int_\x \Gamuuimua[\x,\y;\phibf[\A],\A]  \frac{\delta\phi_i[\x;\A]}{\delta\Anub(\y')}  .
\label{app_c1}
\end{multline} 
To compute $\delta\phibf[\A]/\delta\A$, we take the functional derivative of Eq.~(\ref{phibardef}) wrt $\A$, which gives 
\begin{align}
\frac{\delta\phi_i[\x;\A]}{\delta\Amua(\y)} ={}& - \int_{\x'}  \Gamma^{(2,0)-1}_{ij}[\x,\x';\phibf[\A],\A] \nonumber \\ & \times \Gamuujmua[\x',\y;\phibf[\A],\A] ,
\label{app_c2}
\end{align} 
where $\Gamma^{(2,0)-1}$ is the propagator. From~(\ref{app_c1},\ref{app_c2}), we finally deduce Eq.~(\ref{K2}). 

A similar calculation can be performed in the presence of the regulator term $\Delta S_k[\varphibf,\A]$. The current density then acquires the additional contribution  
\begin{align} 
\DJkmua(\x) &= - \frac{\delta \Delta S_k[\varphibf,\A]}{\delta\Amua(\x)} \nonumber \\ 
&= - \half \int_{\y,\y'} \varphibf(\y) \cdot \frac{\delta \calR_k[\y,\y';\A]}{\delta\Amua(\x)} \varphibf(\y')  
\label{DJmuadef} 
\end{align} 
but the linear response is still given by~(\ref{Kdef}). In the quantum model, one finds that $\Kmunuabk(i\wn)$ is given by $-\Gamkzdmunuab(i\wn,\bar\phibf)$ with the additional term $-\delta_{\mu\nu}(T^a\bar\phibf)\cdot(T^b\bar\phibf) R'_k(0)$. Since the latter vanishes for $k=0$, it can be ignored.

\subsection{$f$-sum rule} 
\label{app_K:subsec_fsum}

Let us consider the system in the absence of gauge field and external source ($\J=\A=0$) so that $\mean{\Jmua}=\mean{\jmua}=0$. After an infinitesimal gauge transformation $O= 1+\theta^a T^a + \calO(\theta^2)$, the gauge field becomes $\Amua{}' = \dmu\theta^a$ [Eq.~(\ref{gaugetransform})]. Since a gauge transformation does not modify the partition function, $\calZ[\A']=\calZ[\A=0]$, the response to the pure gauge field $\A'$ must vanish, i.e., 
\begin{align}
0 = \mean{\Jmua(\x)} &= \int_{\x'} \Kmunuab(\x-\x') \partial_{x'_\nu} \theta^b(\x') \nonumber \\ 
&= - \int_{\x'} [ \partial_{x'_\nu} \Kmunuab(\x-\x') ] \theta^b(\x') ,
\label{app_c9}
\end{align} 
where the last result is obtained by an integration by part. Since Eq.~(\ref{app_c9}) holds for any $\theta^a$, we obtain 
\begin{equation}
\partial_{x'_\nu} \Kmunuab(\x-\x') = 0 . 
\end{equation}
In Fourier space, this yields the ``$f$-sum rule''~(\ref{fsum}) 
relating the paramagnetic current-current correlation function $\Piparaab$ to the diamagnetic term $\mean{(T^a\varphibf) \cdot (T^b\varphibf)}$. 

Equation~(\ref{fsum}) can be generalized to the case of an arbitrary uniform source $\J$ and therefore to an arbitrary uniform field $\phibf$. Let us first consider the partition function $\calZ[\J,\A]$ [Eq.~(\ref{ZJA})] where $\Amu=(\dmu O)O^T$ is a pure gauge field. For an infinitesimal rotation, $\Amua=\dmu\theta^a+\calO(\theta^2)$, so that
\begin{multline}
\ln \calZ[\J,\A] = \ln \calZ[\J,0] \\
+ \half \int_{\x,\x'} \frac{\delta^2\ln \calZ[\J,\A]}{\delta\Amua(\x)\delta\Anub(\x')} \biggl|_{\A=0} \Amua(\x) \Anub(\x')
\end{multline}
to leading order. The term linear in $\Amua$, being proportional to $\mean{\Jmua}_{\A=0}$, vanishes when the source $\J$ is uniform. Using now $\Amua=\dmu\theta^a$ and Eq.~(\ref{Kdef}), we find 
\begin{multline} 
\ln \calZ[\J,\A] = \ln \calZ[\J,0] \\ + \half \int_{\x,\x'} [ \partial_{x_\mu} \partial_{x'_\nu} \Kmunuab(\x,\x',\J)] \theta^a(\x) \theta^b(\x') 
\label{app_c4}
\end{multline} 
after an integration by part. 

Alternatively, one can calculate the $\calO(\theta^2)$ term in $\ln \calZ[\J,\A]$ by first performing the following gauge transformation,
\begin{equation}
\begin{split} 
\varphibf' &= O^T \varphibf , \\ 
\Amu{}' &= O^T \Amu O + (\dmu O^T) O = \dmu (O^TO) = 0 , 
\end{split}
\end{equation}
such that the transformed gauge field vanishes. Since this transformation has unit Jacobian and leaves the action invariant, i.e., $S[\varphibf,\A]=S[\varphibf',\A']\equiv S[\varphibf']$, we obtain 
\begin{equation}
\calZ[\J,\A] = \int\calD[\varphibf']\, e^{-S[\varphibf'] + \int_\x O^T\J\cdot\varphibf'} \equiv \calZ[O^T\J].
\end{equation}
The $\theta$ dependence now comes from the external source
\begin{equation}
O^T\J  = \Bigl(1 - \theta^a T^a + \half T^a T^b \theta^a \theta^b \Bigr) \J + \calO(\theta^3) 
\end{equation}
and the $\calO(\theta^2)$ contribution to $\ln \calZ[\J,\A]$ is given by 
\begin{align} 
& \half \int_\x \frac{\delta \ln \calZ[\J]}{\delta J_i(\x)}\biggl|_{\J(\x)=\J} (T^aT^b\J(\x))_i \theta^a(\x) \theta^b(\x) \nonumber \\ & 
+ \half \int_{\x,\x'} \frac{\delta^2 \ln \calZ[\J]}{\delta J_i(\x) \delta J_j(\x')}\biggl|_{\J(\x)=\J}\nonumber \\ & 
\times (T^a\J(\x))_i (T^b\J(\x'))_j \theta^a(\x) \theta^b(\x') ,
\end{align}
i.e., 
\begin{align}  
& \half \int_\x \phi_i(\x) (T^aT^b\J(\x))_i \theta^a(\x) \theta^b(\x) \nonumber \\ & 
+ \half \int_{\x,\x'} G_{ij}(\x-\x',\J) \nonumber \\ & 
\times (T^a\J(\x))_i (T^b\J(\x'))_j \theta^a(\x) \theta^b(\x') .
\label{app_c5}
\end{align}

Comparing~(\ref{app_c4}) and (\ref{app_c5}) and using $J_i=\partial_{\phi_i}U=\phi_i U'=\phi_iW$ (with $W=U'=\partial_\rho U$), we obtain 
\begin{multline} 
\partial_{x_\mu} \partial_{x'_\nu} \Kmunuab(\x,\x',\phibf) = \half W(\rho) \phibf \cdot \{T^a,T^b\} \phibf \delta(\x-\x') 
\\ 
+ G_{ij}(\x-\x',\phibf) W(\rho)^2 (T^a\phibf)_i (T^b\phibf)_j
\end{multline} 
where $G(\x-\x,\phibf)\equiv G(\x-\x',\J)$ is the propagator for a uniform external source $\J$ (and thus a uniform field $\phibf$) [Eq.~(\ref{Gunif})] and $\{T^a,T^b\}=T^aT^b+T^bT^a$. In Fourier space 
\begin{equation}
\p^2 \Kparaab(\p,\phibf) = (T^a\phibf)\cdot(T^b\phibf) W(\rho)[W(\rho)G_{\rm T}(\p,\rho)-1] . 
\label{app_c6} 
\end{equation}
In the equilibrium state, either $\phibf=0$ (disordered phase) or $W(\rho_0)=U'(\rho_0)=0$ (ordered phase) and we recover the $f$-sum rule~(\ref{fsum}). From the expression~(\ref{Kpara}) of $\Kparaab$, one sees that the $f$-sum rule~(\ref{app_c6}) is a consequence of the Ward identities~(\ref{wi3}).

\section{Ward identities}
\label{app_wi}

The functions $\Psi_A,\Psi_B,\Psi_C,\bar\Psi_B,\bar\Psi_C$ are not independent but related by Ward identities. The effective action $\Gamma[\phi,\A]$ inherits from the symmetries of the microscopic action~(\ref{action2}) and must therefore be invariant in the gauge transformation~(\ref{gaugetransform}): $\Gamma[\phibf,\A]=\Gamma[\phibf',\A']$. For an infinitesimal transformation, i.e., $O=1+\theta^a T^a$ with $\theta^a\to 0$, 
\begin{equation}
\begin{split}
\phibf' &= \phibf + \theta^a T^a \phibf , \\ 
\Amua{}' &= \Amua + \dmu\theta^a + f_{abc} \theta^b \Amuc , 
\end{split}
\end{equation}
where the $f_{abc}$'s are the structure constants of the so($N$) Lie algebra (Appendix~\ref{app_SON}). Gauge invariance then implies
\begin{align}
\int_\x & \Bigl\lbrace \Gamma^{(1,0)}_i[\x;\phibf,\A] (T^a\phibf)_i \theta^a \nonumber \\ &
+ \Gamma^{(0,1)a}_\mu[\x;\phibf,\A] [ \dmu \theta^a + f_{abc} \theta^b \Amuc ] \Bigr\rbrace = 0 
\end{align}
to lowest order in $\theta$. Since this equation holds for arbitrary $\theta^a(\x)$, we deduce 
\begin{align}
&\Gamma^{(1,0)}_i[\x;\phibf,\A] (T^a\phibf)_i - \dmu \Gamma^{(0,1)a}_\mu[\x;\phibf,\A] \nonumber \\ & 
-f_{abc} \Gamma^{(0,1)b}_\mu[\x;\phibf,\A] \Amuc = 0 . 
\label{app_wi1}
\end{align} 
Taking the functional derivative wrt $\phi_j(\x')$ and then setting $\A=0$ and $\phibf$ uniform, we obtain the Ward identity 
\begin{align}
& \Gamma^{(2,0)}_{ij}(\p,\phibf) (T^a\phibf)_i + \frac{1}{\sqrt{V}} \Gamma^{(1,0)}_i(\p=0,\phibf)T^a_{ij} \nonumber \\ & 
- i p_\mu \Gamma^{(1,1)a}_{j\mu}(-\p,\phibf) = 0 . 
\label{wi1} 
\end{align} 
A second Ward identity is obtained from~(\ref{app_wi1}) by taking the functional derivative wrt $\Anud(\x')$, 
\begin{equation}
\Gamma^{(1,1)d}_{i\nu}(\p,\phibf)(T^a\phibf)_i - i p_\mu \Gamma^{(0,2)ad}_{\mu\nu}(\p,\phibf) = 0 ,
\label{wi2} 
\end{equation}
where we have used $\Gamma^{(0,1)b}_{\nu}(\p=0,\phibf)=0$. Together with~(\ref{vert1}), Eqs.~(\ref{wi1},\ref{wi2}) imply~(\ref{wi3}). 

\begin{widetext} 

\section{Computation of $\calR^{(1)}$ and $\calR^{(2)}$} 
\label{app_Rcal} 

To alleviate the notations we do not write explicitly the $k$ index in the following. Writing $\calR^{(1)}$ as
\begin{equation}
\calR^{(1)a}_\mu(\x,\x',\y) = \frac{\delta}{\delta\Amua(\y)} \half \sum_{n=0}^\infty (-1)^n R_n (\D^2)^n \delta(\x-\x') \Bigl|_{\A=0}  + (\x \leftrightarrow \x')^T ,
\end{equation} 
where $R_n$ is defined by $R(u) = \sum_{n=0}^\infty R_n u^n$, and using
\begin{equation}
\frac{\delta \D(\x)^2}{\delta\Amua(\y)} = - T^a \{\delta(\x-\y) , \dmu \} + \{ T^a,\Amu \} \delta(\x-\y) 
\end{equation} 
with $\{\cdot,\cdot\}$ the anticommutator, $\dmu\equiv\partial_{x_\mu}$ and $\nablabf\equiv\nablabf_{\x}$, one finds 
\begin{equation}
\calR^{(1)a}_\mu(\x,\x',\y) = -\half \sum_{n=0}^\infty (-1)^n R_n \sum_{p=0}^{n-1} (\nablabf^2)^p T^a  \{\delta(\x-\y),\dmu \} (\nablabf^2)^{n-1-p} \delta(\x-\x') + (\x \leftrightarrow \x')^T .
\end{equation}
In Fourier space, this gives
\begin{equation}
\calR^{(1)a}_\mu(\p_1,\p_2,\p_3) = \delta_{\p_1+\p_2+\p_3,0} T^a (ip_{1\mu}-ip_{2\mu}) \frac{R(\p_1^2)-R(\p_2^2)}{\p_1^2-\p_2^2} 
\end{equation}
(we define the Fourier transform so as to eliminate any volume factor).

Proceeding similarly for $\calR^{(2)}$, one obtains 
\begin{multline} 
\calR^{(2)ab}_{\mu\nu}(\x,\x',\y,\y') = \half \sum_{n=0}^\infty (-1)^n R_n \sum_{p=0}^{n-1} \Bigl\{ 
\{T^a,T^b\} (\nablabf^2)^p \delta_{\mu\nu} \delta(\x-\y) \delta(\x'-\y) (\nablabf^2)^{n-p-1} \delta(\x-\x') \\
+ \sum_{q=0}^{p-1} (\nablabf^2)^q T^b \{ \delta(\x-\y'),\dnu \} (\nablabf^2)^{p-1-q}
  T^a \{ \delta(\x-\y),\dmu \} (\nablabf^2)^{n-1-p} \delta(\x-\x')  \\
+ \sum_{q=0}^{n-2-p} (\nablabf^2)^p T^a \{ \delta(\x-\y),\dmu \} (\nablabf^2)^{q}
  T^b \{ \delta(\x-\y'),\dnu \} (\nablabf^2)^{n-2-p-q} \delta(\x-\x') \Bigr\} + (\x \leftrightarrow \x')^T
\end{multline} 
and 
\begin{multline} 
\calR^{(2)ab}_{\mu\nu}(\p_1,\p_2,\p_3,\p_4) = \delta_{\p_1+\p_2+\p_3+\p_4,0} 
\Bigl\{ - \delta_{\mu\nu} \{T^a,T^b\} \frac{R(\p_1^2)-R(\p_2^2)}{\p_1^2-\p_2^2}  
- \Bigl[ T^a T^b (2p_{1\mu}+p_{3\mu}) (2p_{2\nu}+p_{4\nu}) \\ \times 
 \frac{R(p_1^2) [(\p_1+\p_3)^2-\p_2^2] + R\left((\p_1+\p_3)^2\right) (\p_2^2-\p_1^2)+R\left(\p_2^2\right) [\p_1^2-(\p_1+\p_3)^2]}{[(\p_1+\p_3)^2-\p_2^2](\p_2^2-\p_1^2) [\p_1^2-(\p_1+\p_3)^2 ]} + (a \leftrightarrow b, \mu \leftrightarrow \nu, \p_3 \leftrightarrow \p_4 ) \Bigr] 
\Bigr\} 
\end{multline}

To order $\p^2$, we finally find
\begin{align} 
\calR^{(1)a}_\mu(\p+\q,-\q,-\p) ={}& i T^a \Bigl\{ 
2q_\mu \Bigl[ R' + \Bigl(\p\cdot\q + \frac{\p^2}{2} \Bigr)R'' + \frac{2}{3} (\p\cdot\q)^2 R''' \Bigr] + p_\mu [ R'+(\p\cdot\q)R''] \Bigr\} \nonumber \\
\calR^{(2)ab}_{\mu\nu}(\q,-\q,\p,-\p) ={}&   - \Bigl\{  
\{T^a, T^b \} \Bigl[ \delta_{\mu \nu} R' + \half p_\mu p_\nu R'' + \frac{2}{3} (q_\mu p_\nu+p_\mu q_\nu) (\p\cdot\q) R'''  \nonumber \\ & 
+ 2 q_\mu q_\nu \Bigl( R'' + \third p^2R'''+\third (\p\cdot\q)^2R'''' \Bigr) \Bigr] 
+ [T^a, T^b] \Bigl[ (q_\mu p_\nu+p_\mu q_\nu)R'' + \frac{2}{3}q_{\mu}q_{\nu} (\p\cdot\q)R''' \Bigr] \Bigr\} , 
\end{align}  
where $R\equiv R(\q^2)$, $R'\equiv \partial_{\q^2} R(\q^2)$, etc. With Eq.~(\ref{rdef}), we obtain 
\begin{equation}
\begin{split} 
\partial^n_{\q^2} R(\q^2) &= Z_k k^{2(1-n)} \bar r^{(n)}(y) , \\
\dt \partial^n_{\q^2} R(\q^2) &= - Z_k k^{2(1-n)} \{ [\eta_k + 2(n-1)] \bar r^{(n)}(y) + 2 y \bar r^{(n+1)}(y) \} ,  
\end{split}
\end{equation}
where $y=\q^2/k^2$, $\bar r(y)=yr(y)$ and $\bar r^{(n)}(y)=\partial^n_y \bar r(y)$.

\section{Flow equations} 
\label{app_rgeq} 

To alleviate the notations we do not write explicitly the $k$ index in the following. 

\subsection{Vertices} 

In addition to $\Gamma^{(2,0)}$, $\Gamma^{(1,1)}$ and $\Gamma^{(0,2)}$ [Eqs.~(\ref{GamAB},\ref{Gam11etc},\ref{Gam02})], the vertices appearing in the flow equations~(\ref{rgeq}) are
\begin{equation}
\begin{split} 
\Gamma^{(3,0)}_{i_1i_2i_3}(\p_1,\p_2,\p_3) ={}& \delta_{i_1i_2} \phi_{i_3} \Bigl( - Z' \p_1\cdot \p_2 + \frac{Y}{2}  \p_3^2 + W' \Bigr) + \phi_{i_1} \phi_{i_2} \phi_{i_3} \Bigl( - \frac{Y'}{2}  \p_2\cdot\p_3 + W'' \Bigr) + \perm(1,2,3) , \\ 
\Gamma^{(4,0)}_{i_1i_2i_3i_4}(\p_1,\p_2,\p_3,\p_4) ={}& \delta_{i_1i_2} \delta_{i_3i_4} \Bigl( -Z' \p_1\cdot\p_2  + \frac{Y}{2} (\p_1+\p_2)^2 +W' \Bigr) + \phi_{i_1} \phi_{i_2} \phi_{i_3} \phi_{i_4} \Bigl( - \frac{Y''}{2} \p_3\cdot\p_4 + W''' \Bigr) \\ &
+ \delta_{i_1i_2} \phi_{i_3} \phi_{i_4} \Bigl( -Z''\p_1\cdot\p_2  - \frac{Y'}{2}  \p_3\cdot\p_4 + \frac{Y'}{2} (\p_1+\p_2)^2 + W'' \Bigr) + \perm(1,2,3,4)  , \\ 
\Gamma^{(2,1)a}_{i_1i_2,\mu}(\p_1,\p_2,\p_3) ={}&  Z' [ (T^a\phibf)_{i_1} \phi_{i_2} i p_{1\mu} + (T^a\phibf)_{i_2} \phi_{i_1} i p_{2\mu} ] + Z T^a_{i_1i_2} (ip_{1\mu}-ip_{2\mu}) , \\ 
\Gamma^{(3,1)a}_{i_1i_2i_3,\mu}(\p_1,\p_2,\p_3,\p_4) ={}& [ Z' \delta_{i_1i_2} + Z'' \phi_{i_1}\phi_{i_2} ] (T^a\phibf)_{i_3} i p_{3\mu} + Z' \phi_{i_1} T^a_{i_2i_3} i(p_{2\mu}-p_{3\mu}) + \perm(1,2,3) ,  \\ 
\Gamma^{(2,2)ab}_{i_1i_2,\mu\nu}(\p_1,\p_2,\p_3,\p_4) ={}& \delta_{\mu\nu} \{ (Z' \delta_{i_1i_2} + Z'' \phi_{i_1}\phi_{i_2} ) (T^a\phibf)\cdot(T^b\phibf) - Z' [\phi_{i_1}(\{T^a,T^b\}\phibf)_{i_2} + \perm(1,2) ] 
- Z \{T^a,T^b\}_{i_1i_2} \} \\ & - (\delta_{\mu\nu}\p_4\cdot\p_3-p_{4\mu} p_{3\nu} ) \{ \delta_{ab} (X_1' \delta_{i_1i_2} + X_1'' \phi_{i_1}\phi_{i_2} )  + (X_2' \delta_{i_1i_2} + X_2'' \phi_{i_1}\phi_{i_2} ) (T^a\phibf)\cdot(T^b\phibf) \\ & - X_2' [ \phi_{i_1}(\{T^a,T^b\}\phibf)_{i_2} + \perm(1,2) ] - X_2 \{T^a,T^b\}_{i_1i_2} \} ,
\end{split} 
\label{vert3}
\end{equation}
where $\perm(1,\cdots,n)$ denotes all (different) terms obtained by permutation of $(\p_1,i_1;\cdots;\p_n,i_n)$. In~(\ref{vert3}) we write neither the Kronecker delta expressing the conservation of total momentum nor the volume factors. The vertices in the lhs are functions of the uniform field $\phibf$; $W$, $Z$, $Y$, $X_1$ and $X_2$ depend on $\rho$, and we use the notation $Z'=\partial_\rho Z$, etc. 

\subsection{Flow equations} 

The flow equations are given by 
\begin{align}
\partial_t W ={}& \frac{(N-1)}{2} \left(W'I_{0|1|0|0}+Z'I_{1|1|0|0}\right) + I_{0|0|1|0} \left( \rho  W''+\frac{3}{2} W'\right) + \frac{1}{2}I_{1|0|1|0} \left(\rho Y'+Y+Z'\right),
   \\
\partial_t Z^{(2,0)} ={}& -\frac{2 \rho  \left(W'\right)^2 I_{1|0,0,1|1|0}}{D}-\frac{2 \rho  \left(W'\right)^2 I_{1|1|0,0,1|0}}{D}-\frac{2 \rho  Y W'
   I_{2|0,0,1|1|0}}{D}
   \nonumber \\ 
   &-\frac{2 \rho  Y W' I_{2|1|0,0,1|0}}{D}-\frac{\rho  W' I_{1|1|0,1|0} \left((D+4) Y-4 Z'\right)}{D}-\frac{\rho  W'
   I_{1|0,1|1|0} \left(D Y+4 Z'\right)}{D}
   \nonumber \\ 
   &-\frac{\rho  Y^2 I_{3|0,0,1|1|0}}{2 D}-\frac{\rho  Y^2 I_{3|1|0,0,1|0}}{2 D}-\frac{\rho 
   I_{1|1|1|0} \left((D+2) Y^2+2 (D-2) Y Z'+4 \left(Z'\right)^2\right)}{2 D}
   \nonumber \\ 
   &-\frac{\rho  Y I_{2|1|0,1|0} \left((D+8) Y-8 Z'\right)}{4 D}-\frac{\rho  Y
   I_{2|0,1|1|0} \left(D Y+8 Z'\right)}{4 D}+\frac{1}{2} I_{0|1|0|0} \left((N-1) Z'+Y\right)
   \nonumber \\ 
   &-\rho  \left(W'\right)^2 I_{0|0,1|1|0}-\rho 
   \left(W'\right)^2 I_{0|1|0,1|0}-\rho  W' I_{0|1|1|0} \left(Y+2 Z'\right)+I_{0|0|1|0} \left(\rho  Z''+\frac{Z'}{2}\right), \label{dZ20} 
   \\
\partial_t Z^{(1,1)} ={}&  -\frac{2 W' I_{1|0,1|1|0} \left(\rho  Z'+Z\right)}{D}+\frac{2 W' I_{1|1|0,1|0} \left(\rho  Z'+Z\right)}{D}-\frac{2 W' I_{1|0,1|1|1}}{D}+\frac{2 W'
   I_{1|1|0,1|1}}{D}
   \nonumber \\ 
   &-\frac{Y I_{2|0,1|1|0} \left(\rho  Z'+Z\right)}{D}+\frac{Y I_{2|1|0,1|0} \left(\rho 
   Z'+Z\right)}{D}+\frac{I_{1|1|1|1} \left(Y-2 Z'\right)}{D}-\frac{Y
   I_{2|0,1|1|1}}{D}
   \nonumber \\ 
   &-\frac{I_{1|1|1|0} \left(Z' ((D-2) \rho  Y+4 Z)-2 Y Z+4 \rho  \left(Z'\right)^2\right)}{2 D}+\frac{Y I_{2|1|0,1|1}}{D}+\frac{1}{2} (N-1) Z' I_{0|1|0|0}
   \nonumber \\ 
   &-\rho  W' Z' I_{0|1|1|0}+I_{0|0|1|0}
   \left(\rho  Z''+\frac{3 Z'}{2}\right), 
\\
\partial_t Z^{(0,2)} ={}&   \frac{4 Z^2 W'I_{1|2|1|0}}{D}+\frac{8 Z W'I_{1|2|1|1}}{D}+\frac{4 W'I_{1|2|1|2}}{D}-\frac{2 W'I_{1|1|1|0,1}}{D}+\frac{2 Y Z^2
  I_{2|2|1|0}}{D}+\frac{2 YI_{2|2|1|2}}{D}
  \nonumber \\ 
  &+I_{1|1|1|0} \left(-\frac{2 \rho  \left(Z'\right)^2}{D}-\frac{4 Z Z'}{D}-\frac{Y Z}{2}\right)+I_{1|1|1|1} \left(-\frac{4
   Z'}{D}-\frac{Y}{2}\right)+\frac{4 Y ZI_{2|2|1|1}}{D}
   \nonumber \\ 
  &-\frac{YI_{2|1|1|0,1}}{D}+I_{0|1|0|0}
   \frac{\left(N -1\right)Z'}{2}-Z W'I_{0|1|1|0}-W'I_{0|1|1|1}+I_{0|0|1|0} \left(\rho  Z''+\frac{5 Z'}{2}\right),
\\
\partial_t Y ={}&   I_{0|0,1|1|0} \left(W'\right)^2+I_{0|1|0,1|0} \left(W'\right)^2-(N-1) I_{0|1,1|0|0} \left(W'\right)^2+\frac{2
   I_{1|0,0,1|1|0} \left(W'\right)^2}{D}
   \nonumber \\ 
  &+\frac{2 I_{1|1|0,0,1|0} \left(W'\right)^2}{D}-\frac{2 (N-1) I_{1|1,0,1|0|0}
   \left(W'\right)^2}{D}-(N-1) Y I_{0|2|0|0} W'
   \nonumber \\ 
  &+\frac{2 Y I_{2|0,0,1|1|0} W'}{D}+\frac{2 Y I_{2|1|0,0,1|0}
   W'}{D}+\frac{I_{1|1|0,1|0} \left((D+4) Y-4 Z'\right) W'}{D}
   \nonumber \\ 
  &-\frac{2 (D+2) (N-1) I_{1|1,1|0|0} Z' W'}{D}-\frac{4 (N-1)
   I_{2|1,0,1|0|0} Z' W'}{D}+I_{0|1|1|0} \left(Y+2 Z'\right) W'
   \nonumber \\ 
  &+\frac{I_{1|0,1|1|0} \left(D Y+4 Z'\right) W'}{D}-\frac{(D+4) (N-1)
   I_{2|1,1|0|0} \left(Z'\right)^2}{D}-I_{0|0|1,1|0} \left(3 W'+2 \rho  W''\right)^2
   \nonumber \\ 
  &-\frac{(2 D+1) I_{1|0|2|0} \left(Y+\rho  Y'+Z'\right)^2}{D}-\frac{(D+4) I_{2|0|1,1|0} \left(Y+\rho 
   Y'+Z'\right)^2}{D}
   \nonumber \\ 
  &-\frac{2 I_{1|0|1,0,1|0} \left(3 W'+2 \rho  W''\right)^2}{D}+\frac{I_{0|1|0|0}
   \left((N-1) \rho  Y'-Y\right)}{2 \rho }+\frac{1}{4} Y I_{2|0,1|1|0} \left(Y+\frac{8 Z'}{D}\right)
   \nonumber \\ 
  &+\frac{Y I_{2|1|0,1|0} \left((D+8) Y-8 Z'\right)}{4 D}-\frac{(N-1) I_{1|2|0|0} Z' \left(D
   Y+Z'\right)}{D}
   \nonumber \\ 
  &+\frac{I_{1|1|1|0} \left((D+2) Y^2+2 (D-2) Z' Y+4 \left(Z'\right)^2\right)}{2
   D}+\frac{I_{3|0,0,1|1|0}}{2D} Y^2+\frac{I_{3|1|0,0,1|0}}{2D} Y^2
   \nonumber \\ 
  & -\frac{2\left (N-1\right )I_{3|1,0,1|0|0}}{D}  \left(Z'\right)^2-\frac{2 I_{3|0|1,0,1|0} \left(Y+\rho  Y'+Z'\right)^2}{D}
   \nonumber \\ 
  &-2 I_{0|0|2|0} \left(Y+\rho  Y'+Z'\right) \left(3 W'+2 \rho  W''\right)
   \nonumber \\ 
  &-\frac{2 (D+2)
   I_{1|0|1,1|0} \left(Y+\rho  Y'+Z'\right) \left(3 W'+2 \rho  W''\right)}{D}
   \nonumber \\ 
  &-\frac{4 I_{2|0|1,0,1|0} \left(Y+\rho  Y'+Z'\right) \left(3 W'+2 \rho 
   W''\right)}{D}+I_{0|0|1|0} \left(\frac{Y}{2 \rho }+\frac{5 Y'}{2}+\rho  Y''\right),
   \\
   \partial_t X_1 ={}&-\frac{4 Z^2 I_{1|1,1|0|0}}{D}-\frac{8 Z^2 I_{2|1,0,1|0|0}}{D (D+2)}-\frac{4 Z I_{1|2|0|0,1}}{D}-\frac{8 Z I_{1|1,1|0|1}}{D}-\frac{16
   Z I_{2|2|0|0,0,1}}{3 D (D+2)}
   \nonumber \\ 
   &-\frac{16 Z I_{2|1,1|0|0,1}}{D (D+2)}-\frac{16 Z I_{2|1,0,1|0|1}}{D (D+2)}-\frac{4
   I_{1|2|0|1,1}}{D}-\frac{4 I_{2|2|0|0,2}}{D (D+2)}-\frac{4 I_{1|1,1|0|2}}{D}
   \nonumber \\ 
   &+\frac{4 I_{1|1|0|0,0,1}}{3 D}-\frac{16
   I_{2|2|0|1,0,1}}{3 D (D+2)}-\frac{16 I_{2|1,1|0|1,1}}{D (D+2)}+\frac{4 I_{2|1|0|0,0,0,1}}{3 D (D+2)}-\frac{8 I_{2|1,0,1|0|2}}{D
   (D+2)}
   \nonumber \\ 
   &+\frac{1}{2} I_{0|1|0|0} \left((N-1) X_1'+4 X_2\right)+I_{0|0|1|0} \left(\rho  X_1''+\frac{X_1'}{2}\right),
   			\label{dX1app} \\
   \partial_t X_2 ={}& \frac{2 I_{1|1,1|0|0} Z^2}{D \rho }+\frac{4 I_{2|1,0,1|0|0} Z^2}{D(D+2)\rho}+\frac{2 I_{1|2|0|0,1} Z}{D \rho }+\frac{4
   I_{1|1,1|0|1} Z}{D \rho }+\frac{8 I_{2|2|0|0,0,1} Z}{3D(D+2)\rho}
   \nonumber \\ 
   &+\frac{8 I_{2|1,1|0|0,1} Z}{D(D+2)\rho}+\frac{8
   I_{2|1,0,1|0|1} Z}{D(D+2)\rho}-\frac{I_{1|0,1|1|0} \left(Z+\rho  Z'\right)^2}{D \rho }-\frac{I_{1|1|0,1|0} \left(Z+\rho  Z'\right)^2}{D
   \rho }
   \nonumber \\ 
   &-\frac{2 I_{2|0,0,1|1|0} \left(Z+\rho  Z'\right)^2}{D (D+2) \rho }-\frac{2 I_{2|1|0,0,1|0} \left(Z+\rho  Z'\right)^2}{D (D+2) \rho }-\frac{2
   I_{1|1|1|1,1}}{D \rho }+\frac{2 I_{1|2|0|1,1}}{D \rho }
   \nonumber \\ 
   &-\frac{2 I_{2|1|1|0,2}}{D(D+2)\rho}+\frac{2 I_{2|2|0|0,2}}{D(D+2)\rho}-\frac{I_{1|0,1|1|2}}{D \rho }+\frac{I_{1|0|1|0,0,1}}{3 D \rho }-\frac{I_{1|1|0,1|2}}{D \rho }+\frac{2 I_{1|1,1|0|2}}{D \rho
   }
   \nonumber \\ 
   &-\frac{I_{1|1|0|0,0,1}}{3 D \rho }-\frac{8 I_{2|1|1|1,0,1}}{3D(D+2)\rho}+\frac{8 I_{2|2|0|1,0,1}}{3D(D+2)\rho}-\frac{4
   I_{2|0,1|1|1,1}}{D(D+2)\rho}+\frac{I_{2|0|1|0,0,0,1}}{3D(D+2)\rho}-\frac{4 I_{2|1|0,1|1,1}}{D(D+2)\rho}
   \nonumber \\ 
   &+\frac{8
   I_{2|1,1|0|1,1}}{D(D+2)\rho}-\frac{I_{2|1|0|0,0,0,1}}{3D(D+2)\rho}-\frac{2 I_{2|0,0,1|1|2}}{D(D+2)\rho}-\frac{2
   I_{2|1|0,0,1|2}}{D(D+2)\rho}+\frac{4 I_{2|1,0,1|0|2}}{D(D+2)\rho}
   \nonumber \\ 
   &+\frac{I_{0|1|0|0} \left((N-1) \rho 
   X_2'-X_2\right)}{2 \rho }-\frac{2 I_{1|1|1|0,1} \left(Z+\rho  Z'\right)}{D \rho }-\frac{2 I_{1|0,1|1|1} \left(Z+\rho  Z'\right)}{D \rho }
   \nonumber \\ 
   &-\frac{2
   I_{1|1|0,1|1} \left(Z+\rho  Z'\right)}{D \rho }-\frac{8 I_{2|1|1|0,0,1} \left(Z+\rho  Z'\right)}{3 D (D+2) \rho }-\frac{4 I_{2|0,1|1|0,1}
   \left(Z+\rho  Z'\right)}{D (D+2) \rho }
   \nonumber \\ 
   &-\frac{4 I_{2|1|0,1|0,1} \left(Z+\rho  Z'\right)}{D (D+2) \rho }-\frac{4 I_{2|0,0,1|1|1} \left(Z+\rho  Z'\right)}{D (D+2)
   \rho }-\frac{4 I_{2|1|0,0,1|1} \left(Z+\rho  Z'\right)}{D (D+2) \rho }
   \nonumber \\ 
   &+I_{0|0|1|0} \left(\frac{X_2}{2 \rho }+\frac{5 X_2'}{2}+\rho  X_2'' \right) ,
\end{align}
where $D=d+1$. $\dt Z^{(2,0)}$, $\dt Z^{(1,1)}$ and $\dt Z^{(0,2)}$ correspond to the equation $\dt Z$ obtained from the vertices $\Gamma^{(2,0)}$, $\Gamma^{(1,1)}$ and $\Gamma^{(0,2)}$, respectively. We have introduced the ``threshold'' function
\begin{equation}
I_{n|\alpha_i|\beta_i|\gamma_i}(\rho) = \tdt \int_\q (\q^2)^n \prod_i (G_{k,\rm T}^{(i)}(\q,\rho))^{\alpha_i}(G_{k,\rm L}^{(i)}(\q,\rho))^{\beta_i}(R^{(i)}_k(\q))^{\gamma_i} , 
\label{Idef} 
\end{equation}
e.g. 
\begin{equation}
\begin{split} 
I_{1|1|1|0} &= \tdt \int_\q \q^2 G_{k,\rm T}(\q,\rho) G_{k,\rm L}(\q,\rho) , \\  
I_{2|2|0,0,1|0,1} &= \tdt \int_\q (\q^2)^2 (G_{k,\rm T}(\q,\rho))^2 G_{k,\rm L}^{(2)}(\q,\rho) R'_k(\q) ,
\end{split}
\end{equation}
where $G^{(n)}=\partial^n_{\q^2} G$ and $\tdt=k\tdk$. The initial conditions are determined by $\Gamma_{\Lambda}=S$, which yields
\begin{align}
W_\Lambda(\rho)&=r_0+\dfrac{u_0}{3N}\rho,&Z_\Lambda(\rho)&=1,&Y_{\Lambda}(\rho)&=X_{1,\Lambda}(\rho)=X_{2,\Lambda}(\rho)=0.
\end{align}
In practice, we solve the flow equations using dimensionless variables (see Sec.~\ref{sec_num}).

\section{Large-$N$ limit: computation of $I_{X_1}$} 
\label{app_largeN_IX1} 

From Eq.~(\ref{dX1app}) we deduce that the $\calO(1)$ contribution to $\dt X_{1,k}$ in the large-$N$ limit is given by~(\ref{dX1}) with 
\begin{align}
I_{X_1}={}&\frac{1}{6 D (D+2)} \int_\q \bigl( -24 D q^2 R_k' R_k'' G_{k,\rm T}^2-24 D q^2 R_k'{}^2 G_{k,\rm T} G_{k,\rm T}'-48 D q^2 R_k' G_{k,\rm T} G_{k,\rm T}'-24 D q^2 R_k'' G_{k,\rm T}^2
\nonumber \\ 
&+8 D q^2 R_k^{(3)} G_{k,\rm T}-24 D q^2 G_{k,\rm T} G_{k,\rm T}'-96 q^4
   R_k'' G_{k,\rm T} G_{k,\rm T}'-96 q^4 R_k' R_k'' G_{k,\rm T} G_{k,\rm T}'-48 q^4 R_k'{}^2 G_{k,\rm T} G_{k,\rm T}''
   \nonumber \\ 
   &-96 q^4 R_k' G_{k,\rm T} G_{k,\rm T}''-32 q^4 R_k^{(3)} R_k' G_{k,\rm T}^2-24 q^4
   R_k''{}^2 G_{k,\rm T}^2-32 q^4 R_k^{(3)} G_{k,\rm T}^2+8 q^4 R_k^{(4)} G_{k,\rm T}
   \nonumber \\ 
   &-48 q^4 G_{k,\rm T} G_{k,\rm T}''-48 q^2 R_k' R_k'' G_{k,\rm T}^2-48 q^2 R_k'{}^2 G_{k,\rm T} G_{k,\rm T}'-96 q^2
   R_k' G_{k,\rm T} G_{k,\rm T}'-48 q^2 R_k'' G_{k,\rm T}^2
  \nonumber \\ 
   &+16 q^2 R_k^{(3)} G_{k,\rm T}-48 q^2 G_{k,\rm T} G_{k,\rm T}' \bigl) .
\label{app_IX1_1}
\end{align}
In the above integral the primes denote derivatives wrt $q^2=\q^2$, and we omit the dependence on $q^2$ of the functions for the sake of shortness. Using $R'_k=-1-G_{k,\rm T}'/G_{k,\rm T}^2$, one can remove $R_k'$ and its derivatives from~(\ref{app_IX1_1}) and obtain the simpler expression 
\begin{align}
I_{X_1}={}& - \int_\q \frac{4 q^2}{3 D (D+2) G_{k,\rm T}^4}\bigl\{ 3 G_{k,\rm T} G_{k,\rm T}'{}^2 [(D+2) G_{k,\rm T}'+6 q^2 G_{k,\rm T}''] \nonumber \\ &
-G_{k,\rm T}^2 \bigl(G_{k,\rm T}' [3 (D+2) G_{k,\rm T}''+4 q^2
   G_{k,\rm T}^{(3)}]+3 q^2 G_{k,\rm T}''{}^2 \bigr) 
   +G_{k,\rm T}^3 [(D+2) G_{k,\rm T}^{(3)}+q^2 G_{k,\rm T}^{(4)}] - 12 q^2 G_{k,\rm T}'{}^4 \bigr\}.
\end{align} 
Integrating by parts yields~(\ref{IX1}).

\section{One-loop approximation} 
\label{app_1loop} 

\subsection{$\Gamma^{(2,0)}_{ij}$, $\Gamuuimua$ and $\Gamzdmunuab$} 

From~(\ref{Gam1loop}) we deduce 
\begin{equation}
\begin{split}
\Gamma^{(2,0)}_{ij}(\x,\x') ={}& S^{(2,0)}_{ij}(\x,\x') +  \half \Tr [ (G-G_\Lamb) S^{(4,0)}_{ij}(\x,\x')
- G S^{(3,0)}_{i}(\x) G S^{(3,0)}_{i}(\x')
+ G_\Lamb S^{(3,0)}_{i}(\x) G_\Lamb S^{(3,0)}_{i}(\x') ] , \\ 
\Gamuuimua(\x,\y) ={}& \Suuimua(\x,\y)  
- \half \Tr \{ G S^{(3,0)}_i(\x) G S^{(2,1)a}_\mu(\y) - G_\Lamb S^{(3,0)}_i(\x) G_\Lamb [ S^{(2,1)a}_\mu(\y) + \calR^{(1)a}_{\Lamb,\mu}(\y) ]
 \} , \\ 
\Gamzdmunuab(\y,\y') ={}& \Szdmunuab(\y,\y')+ \half \Tr \{ G \Sddmunuab(\y,\y') 
- G_\Lamb [ \Sddmunuab(\y,\y') + \calR^{(2)ab}_{\Lamb,\mu\nu}(\y,\y') ] \\ & 
- G S^{(2,1)a}_\mu(\y) G S^{(2,1)b}_\nu(\y') 
+ G_\Lamb [ S^{(2,1)a}_\mu(\y) + \calR^{(1)a}_{\Lamb,\mu}(\y) ] G_\Lamb [ S^{(2,1)b}_\nu(\y') + \calR^{(1)b}_{\Lamb,\nu}(\y') ] \} ,
\end{split}
\label{Gam1loop2}
\end{equation}
where $G\equiv G_{k=0}$ and, for clarity, we do not write the dependence on the uniform field $\phibf$.
$S^{(4,0)}$, $S^{(3,0)}$, $S^{(1,1)}$, $S^{(2,1)}$and $S^{(2,2)}$ are easily deduced from Eqs.~(\ref{Gam11etc},\ref{vert3}); $\calR^{(1)a}_\mu$ and $\calR^{(2)ab}_{\mu\nu}$ are given in Appendix~\ref{app_Rcal}.

\subsection{$Z(\rho)$} 
\label{app_1loop:subsec_Z} 

From Eqs.~(\ref{Gam1loop2}), we obtain three expressions of $Z(\rho)$ to one-loop order, 
\begin{align}
Z(\rho) ={}& 1-\rho W'_\Lamb{}^2 \int_\q \Bigl[ G'_{\rm T} G_{\rm L} + G_{\rm T} G'_{\rm L} - G'_{\Lamb,\rm T} G_{\Lamb,\rm L} - G_{\Lamb,\rm T} G'_{\Lamb,\rm L} \nonumber \\ &  
+ \frac{2}{D} q^2 (   G''_{\rm T} G_{\rm L} + G_{\rm T} G''_{\rm L} 
- G''_{\Lamb,\rm T} G_{\Lamb,\rm L} - G_{\Lamb,\rm T} G''_{\Lamb,\rm L} ) \Bigr] ,
\label{Z20oneloop} \\ 
Z(\rho) ={}& 1 - \frac{2}{D} W'_\Lamb \int_\q q^2 [ (1+R'_\Lamb) ( G_{T,\Lamb} G'_{L,\Lamb}  
- G'_{T,\Lamb} G_{L,\Lamb} )  - G_{T} G'_{L}  + G'_{T} G_{L} ] , 
\label{Z11oneloop} \\
Z(\rho) ={}& 1 + \frac{1}{2\rho} \int_\q \Bigl[ G_{\rm L}-G_{\rm T} - (G_{\Lamb,\rm L}-G_{\Lamb,\rm T}) \Bigl( 1 + R' + \frac{2}{D} q^2 R'' \Bigr) \nonumber \\ & 
- \frac{4}{D} q^2 G_{\rm T} (G_{\rm L}-G_{\rm T}) 
+ \frac{4}{D} (1+R')^2 q^2 G_{T,\Lamb}(G_{\Lamb,\rm L}-G_{\Lamb,\rm T}) \Bigr]
\label{Z02oneloop} 
\end{align}  

Let us show that Eqs.~(\ref{Z20oneloop}) and (\ref{Z11oneloop}) agree. The difference between the two expressions gives 
\begin{align} 
& - \frac{2}{D} \rho W'_\Lamb{}^2 \int_\q q^2 ( G''_{\rm T} G_{\rm L} + G_{\rm T} G''_{\rm L} - G''_{\Lamb,\rm T} G_{\Lamb,\rm L} - G_{\Lamb,\rm T} G''_{\Lamb,\rm L} )  
- \rho W'_\Lamb{}^2 \int_\q ( G'_{\rm T} G_{\rm L} + G_{\rm T} G'_{\rm L} - G'_{\Lamb,\rm T} G_{\Lamb,\rm L} - G_{\Lamb,\rm T} G'_{\Lamb,\rm L} ) \nonumber \\ &
+ \frac{2}{D} W'_\Lamb \int_\q q^2 [  (1+R') (G_{\Lamb,\rm T} G'_{\Lamb,\rm L} - G'_{\Lamb,\rm T} G_{\Lamb,\rm L} ) 
- G_{\rm T} G'_{\rm L} + G'_{\Lamb,\rm T} G_{\Lamb,\rm L} ] .
\label{app_1loop_1}
\end{align}
Using $G'_{\Lamb,\rm T}=-(1+R'_\Lamb)G_{\Lamb,\rm T}^2$ and $G''_{\Lamb,\rm T}=-R''_\Lamb G_{\Lamb,\rm T}^2+2(1+R'_\Lamb)^2G_{\Lamb,\rm T}^3$ (and similar relations for $G_{\Lamb,\rm L}$), as well as $G_{\Lamb,\rm L}-G_{\Lamb,\rm T}=2\rho W'_\Lamb G_{\Lamb,\rm T} G_{\Lamb,\rm L}$, we obtain 
\begin{equation}
\frac{4}{D} \rho W'_\Lamb{}^2 \int_\q q^2(1+R'_\Lamb)^2 ( G_{\Lamb,\rm T}^2 G_{\Lamb,\rm L}^2 + G_{\Lamb,\rm T}^3 G_{\Lamb,\rm L} + G_{\Lamb,\rm T} G_{\Lamb,\rm L}^3 )
- \rho W'_\Lamb{}^2 \int_\q \Bigl( 1+R_\Lamb'+\frac{2}{D}q^2 R''_\Lamb \Bigr) ( G_{\Lamb,\rm T}^2 G_{\Lamb,\rm L} + G_{\Lamb,\rm T} G^2_{\Lamb,\rm L} ) 
\label{app_1loop_2}
\end{equation}
for the part of~(\ref{app_1loop_1}) which depends on $G_\Lamb$. Noting that 
\begin{equation}
\partial_{q^2} (G_{\Lamb,\rm T}^2 G_{\Lamb,\rm L} + G_{\Lamb,\rm T} G^2_{\Lamb,\rm L} ) = - 2(1+R'_\Lamb) ( G_{\Lamb,\rm T}^2 G_{\Lamb,\rm L}^2 + G_{\Lamb,\rm T}^3 G_{\Lamb,\rm L} + G_{\Lamb,\rm T} G_{\Lamb,\rm L}^3 ) , 
\end{equation}
we can integrate~(\ref{app_1loop_2}) by part (taking $q^2$ as the integration variable), which gives 
\begin{equation}
- 4\frac{v_D}{D} \rho W'_\Lamb{}^2 q^D (1+R'_\Lamb)(G_{\Lamb,\rm T}^2 G_{\Lamb,\rm L} + G_{\Lamb,\rm T} G^2_{\Lamb,\rm L} ) \Bigr|_0^\infty .
\end{equation}
Adding now the part that depends on $G\equiv G_{k=0}$, we obtain 
\begin{equation}
- 4\frac{v_D}{D} \rho W'_\Lamb{}^2 q^D [ (1+R'_\Lamb)(G_{\Lamb,\rm T}^2 G_{\Lamb,\rm L} + G_{\Lamb,\rm T} G^2_{\Lamb,\rm L} ) - G_{\rm T}^2 G_{\rm L} - G_{\rm T} G^2_{\rm L} ] \Bigr|_0^\infty = 0 
\label{app_1loop_3}
\end{equation}
for the difference between the two expressions of $Z(\rho)$ given by Eqs.~(\ref{Z20oneloop}) and (\ref{Z11oneloop}). The equivalence between Eqs.~(\ref{Z20oneloop}) and (\ref{Z02oneloop}) can be shown in a similar way. 
\end{widetext}



\begin{thebibliography}{65}
\bibitem [{no1()}]{no1}%
  \BibitemOpen
  \href@noop {} {}\bibinfo {note} {Lorentz invariance implies that the velocity
  $c$ is not renormalized at zero temperature.}\BibitemShut {Stop}%
\bibitem [{no2()}]{no2}%
  \BibitemOpen
  \href@noop {} {}\bibinfo {note} {Since $U(\rho)$ is a convex function (in
  agreement with the convexity of the Legendre transform $\Gamma$), $U(\rho)$
  is actually constant for $0\leq\rho\leq\rho_0$ in the broken-symmetry
  phase.}\BibitemShut {Stop}%
\bibitem [{no6()}]{no6}%
  \BibitemOpen
  \href@noop {} {}\bibinfo {note} {The Josephson scale $k_J\sim\xi_J^{-1}$ is
  defined by the dimensionless mass of the longitudinal mode being of order
  one, i.e., $2\trho_{0,k_J}\tilde W'_{k_J}(\trho_{0,k_J})\sim
  1$.\cite{Rancon13a}}\BibitemShut {NoStop}%
\end{thebibliography}

%

\end{document}